%% file: main.tex
\documentclass[
    submission,
    UKenglish,
    copyright,
    creativecommons
]{eptcs}
\providecommand{\event}{QPL 2023}

\input{preamble/packages}
\input{preamble/symbols}

\input{preamble/qudit-ZX.tikzstyles}

\title{The Qupit Stabiliser ZX-travaganza: \\Simplified Axioms, Normal Forms and Graph-Theoretic Simplification}
\def\titlerunning{The Qupit Stabiliser ZX-travaganza}

\author{Boldizsár Poór
\institute{University of Oxford}
\institute{Quantinuum, 17 Beaumont Street\\Oxford, OX1 2NA, United Kingdom}
\email{boldizsar.poor@quantinuum.com} \and
Robert I. Booth
\institute{University of Edinburgh}
\email{robert.booth@ed.ac.uk} \and
Titouan Carette
\institute{Centre for Quantum Computer Science,\\ Faculty of Computing, University of Latvia,\\ Raina 19, Riga, Latvia, LV-1586}
\email{titouan.carette@lu.lv} \and
John van de Wetering
\institute{University of Amsterdam}
\email{john@vdwetering.name} \and
Lia Yeh
\institute{University of Oxford}
\institute{Quantinuum, 17 Beaumont Street\\Oxford, OX1 2NA, United Kingdom}
\email{lia.yeh@cs.ox.ac.uk} 
}
\def\authorrunning{Poór et al.}

\begin{document}

\maketitle

\begin{abstract}
    We present a smorgasbord of results on the stabiliser ZX-calculus for odd prime-dimensional qudits (i.e.\@ \emph{qupits}).
    We derive a simplified rule set that closely resembles the original rules of qubit ZX-calculus.
    Using these rules, we demonstrate analogues of the spider-removing local complementation and pivoting rules.
    This allows for efficient reduction of diagrams to the \emph{affine with phases} normal form.
    We also demonstrate a reduction to a unique form, providing an alternative and simpler proof of completeness.
    Furthermore, we introduce a different reduction to the \emph{graph state with local Cliffords} normal form,
    which leads to a novel layered decomposition for qupit Clifford unitaries.
    Additionally, we propose a new approach to handle scalars formally, closely reflecting their practical usage.
    Finally, we have implemented many of these findings in \texttt{DiZX}, a new open-source Python library for qudit ZX-diagrammatic reasoning.
\end{abstract}

\subfile{introduction}

\subfile{qupit-clifford-zx}

\subfile{ap-form}

\subfile{conclusion}

\textbf{Acknowledgements}:
We would like to thank Razin A.~Shaikh for his contributions to the development of \texttt{DiZX}.
LY is supported by an Oxford - Basil Reeve Graduate Scholarship at Oriel College with the Clarendon Fund.
Some of this work was done while BP was a student at the University of Oxford.
The results of~\crefrange{subsec:graph-simplifications}{subsec:ap-form}, \cref{lem:state_colour_change}, and the explicit scalars are also presented in his Master's thesis~\cite{poorUniqueNormalForm2022}.
TC was supported by the ERDF project 1.1.1.5/18/A/020 ``Quantum algorithms: from complexity theory to experiment''.

\bibliographystyle{eptcs}
\bibliography{preamble/references}
    
\appendix

\section*{Appendix}
\subfile{app-minimality}

\subfile{app-lemmas}

\end{document}

%% file: preamble/packages.tex
\usepackage[a-3b, mathxmp]{pdfx} 
\usepackage{xspace}
\usepackage{lmodern}

\usepackage{csquotes}

\usepackage{import}
\usepackage{subfiles}

\usepackage{mathtools}
\usepackage{physics}
\usepackage{amsfonts, amssymb, amsthm}
\usepackage{thmtools}
\usepackage{mathdots}   
\usepackage{mathrsfs}   
\usepackage{stmaryrd}   
\SetSymbolFont{stmry}{bold}{U}{stmry}{m}{n}

\usepackage{tikz}
\usetikzlibrary{decorations.markings}
\usetikzlibrary{quantikz}
\usepackage{tikzit}
\usetikzlibrary{cd}
\definecolor{zx_grey}{RGB}{211,211,211}
\definecolor{zx_red}{RGB}{232,165,165}
\definecolor{zx_green}{RGB}{216,248,216}
\definecolor{dark-gray}{gray}{0.40}

\usepackage{bookmark} 
\hypersetup{
    colorlinks=true,
    citecolor=gray,
	linkcolor=blue,
}

\usepackage{graphicx}
\graphicspath{{figures/}}

\usepackage[all]{hypcap}

\usepackage[capitalise, noabbrev]{cleveref}

%% file: preamble/symbols.tex
\theoremstyle{definition}
\newtheorem{theorem}{Theorem}
\newtheorem{definition}[theorem]{Definition}
\newtheorem{proposition}[theorem]{Proposition}
\newtheorem{lemma}[theorem]{Lemma}

\declaretheoremstyle[
    headfont=\normalfont\bfseries\color{dark-gray},
    bodyfont=\normalfont,
    notefont=\normalfont\bfseries\color{dark-gray},
    notebraces={}{},
    headpunct={.},
    qed=\qedsymbol,
    mdframed={
    linewidth=1.5,
    linecolor=gray,
    hidealllines=true,
    leftline=true,
    skipabove=0,
    innerleftmargin=2mm,
    innerrightmargin=0,
    innertopmargin=0,
    innerbottommargin=.7mm
    }
]{line_proof}

\declaretheorem[name=Proof,style=line_proof,numbered=no]{lproof}

\newcommand{\N}{\mathbb{N}}
\newcommand{\Z}{\mathbb{Z}}

\ifdef{\C}
  {\renewcommand{\C}{\mathbb{C}}}
  {\newcommand{\C}{\mathbb{C}}}

\newcommand{\minu}{\texttt{-}}
\newcommand{\plus}{\texttt{+}}

\newcommand{\interp}[1]{\left\llbracket #1 \right\rrbracket}

\newcommand{\ZXp}{\mathsf{ZX}_p^{\mathrm{Stab}}}

\newcommand{\ZXeq}{\ZXp}

\newcommand{\Fusion}{\tiny{(\hyperref[fig:axioms]{\textsc{Fusion}})}}
\newcommand{\Colour}{\tiny{(\hyperref[fig:axioms]{\textsc{Colour}})}}

\newcommand{\Copy}{\tiny{(\hyperref[fig:axioms]{\textsc{Copy}})}}
\newcommand{\CopyOG}{\tikzlemref{lem:copy_og}}
\newcommand{\Bigebra}{\tiny{(\hyperref[fig:axioms]{\textsc{Bigebra}})}}
\newcommand{\ZElim}{\tikzlemref{lem:green_trivial}}
\newcommand{\XElim}{\tikzlemref{lem:antipode_inverse}}

\newcommand{\Euler}{\tiny{(\hyperref[fig:axioms]{\textsc{Euler}})}}
\newcommand{\One}{\tiny{(\hyperref[fig:axioms]{\textsc{One}})}}
\newcommand{\Gauss}{\tiny{(\hyperref[fig:axioms]{\textsc{Gauss}})}}
\newcommand{\MElim}{\tiny{(\hyperref[fig:axioms]{\textsc{M-Elim}})}}
\newcommand{\Prod}{\tiny{(\hyperref[fig:axioms]{\textsc{Prod}})}}
\newcommand{\Omeg}{\tiny{(\hyperref[fig:axioms]{\textsc{Omega}})}}

\newcommand{\Special}{\tiny{(\hyperref[fig:axioms]{\textsc{Special}})}}

\newcommand{\TextGauss}{\hyperref[fig:axioms]{\textsc{Gauss}}\xspace}
\newcommand{\TextFusion}{\hyperref[fig:axioms]{\textsc{Fusion}}\xspace}

\newcommand{\TextColour}{\hyperref[fig:axioms]{\textsc{Colour}}\xspace}
\newcommand{\TextEuler}{\hyperref[fig:axioms]{\textsc{Euler}}\xspace}
\newcommand{\TextCopy}{\hyperref[fig:axioms]{\textsc{Copy}}\xspace}
\newcommand{\TextMElim}{\hyperref[fig:axioms]{\textsc{M-Elim}}\xspace}
\newcommand{\TextOmega}{\hyperref[fig:axioms]{\textsc{Omega}}\xspace}
\newcommand{\TextProd}{\hyperref[fig:axioms]{\textsc{Prod}}\xspace}
\newcommand{\TextOne}{\hyperref[fig:axioms]{\textsc{One}}\xspace}
\newcommand{\TextZero}{\hyperref[fig:axioms]{\textsc{Zero}}\xspace}

\newcommand{\TextNul}{\hyperref[fig:axioms]{\textsc{Nul}}\xspace}

\newcommand{\TextBigebra}{\hyperref[fig:axioms]{\textsc{Bigebra}}\xspace}
\newcommand{\TextSpecial}{\hyperref[fig:axioms]{\textsc{Special}}\xspace}

\newcommand{\tikzeqref}[1]{\tiny{\textsc{(Eq~\ref{#1})}}}
\newcommand{\tikzlemref}[1]{\tiny{\textsc{(Lem~\ref{#1})}}}
\newcommand{\tikzpropref}[1]{\tiny{\textsc{(Prop~\ref{#1})}}}

\newcommand{\booth}[1]{Lemma #1 of Ref.~\cite{boothCompleteZXcalculiStabiliser2022v3}}

%% file: preamble/qudit-ZX.tikzstyles
\tikzstyle{gn}=[font={\scriptsize\boldmath}, inner sep=1mm, outer sep=-1.8mm, scale=0.8, tikzit shape=circle, draw=black, fill={zx_green}, tikzit draw=black, tikzit fill=green, shape=circle, tikzit category=ZX]
\tikzstyle{rn}=[font={\scriptsize\boldmath}, inner sep=1mm, outer sep=-1.8mm, scale=0.8, tikzit shape=circle, draw=black, fill={zx_red}, tikzit fill=red, tikzit draw=black, shape=circle, tikzit category=ZX]
\tikzstyle{gnphase}=[rounded rectangle, minimum height=11pt, font={\scriptsize\boldmath}, inner sep=1mm, outer sep=-1.8mm, scale=0.8, draw=black, fill={zx_green}, tikzit draw=black, tikzit fill=green, tikzit category=ZX]
\tikzstyle{rnphase}=[rounded rectangle, minimum height=11pt, font={\scriptsize\boldmath}, inner sep=1mm, outer sep=-1.8mm, scale=0.8, draw=black, fill={zx_red}, tikzit fill=red, tikzit draw=black, tikzit category=ZX]
\tikzstyle{had}=[fill={rgb,255: red,255; green,238; blue,2}, draw=black, shape=rectangle, tikzit category=ZX, tikzit draw=black, minimum size=5pt, inner sep=1.5pt, font={\scriptsize\boldmath}, tikzit fill={rgb,255: red,255; green,238; blue,2}]
\tikzstyle{scalar}=[rounded rectangle, rounded rectangle arc length=120, fill=gray, inner sep=2pt, font={\tiny\boldmath}, label distance=1mm, fill opacity=.25, text opacity=1, tikzit category=ZX]
\tikzstyle{gphase}=[rounded rectangle, rounded rectangle arc length=120, fill={zx_green}, inner sep=2pt, font={\tiny\boldmath}, label distance=1mm, fill opacity=.8, text opacity=1, tikzit category=ZX]
\tikzstyle{rphase}=[rounded rectangle, rounded rectangle arc length=120, fill={zx_red}, inner sep=2pt, font={\tiny\boldmath}, label distance=1mm, fill opacity=.6, text opacity=1, tikzit category=ZX]
\tikzstyle{mphase}=[rounded rectangle, rounded rectangle arc length=120, fill=gray, inner sep=2pt, font={\tiny\boldmath}, label distance=1mm, fill opacity=.6, text opacity=1, tikzit category=ZX]
\tikzstyle{lmat}=[shape=signal, signal to=west, signal from=east, fill={zx_grey}, draw=black, minimum height=6pt, inner sep=1pt, font={\scriptsize  \boldmath}, tikzit fill=gray, tikzit category=GLA]
\tikzstyle{rmat}=[shape=signal, signal to=east, signal from=west, fill={zx_grey}, draw=black, minimum height=6pt, inner sep=1pt, font={\scriptsize  \boldmath}, tikzit fill=gray, tikzit category=GLA]
\tikzstyle{dmat}=[shape=signal, signal to=west, signal from=east, fill={zx_grey}, draw=black, minimum height=6pt, inner sep=1pt, font={\scriptsize  \boldmath}, tikzit fill=gray, tikzit category=GLA, rotate=90]
\tikzstyle{umat}=[shape=signal, signal to=east, signal from=west, fill={zx_grey}, draw=black, minimum height=6pt, inner sep=1pt, font={\scriptsize  \boldmath}, tikzit fill=gray, tikzit category=GLA, rotate=90]
\tikzstyle{graph_vertex}=[fill=black, draw=black, shape=circle, tikzit category=mbqc, minimum size=2.4mm, inner sep=.8mm]
\tikzstyle{graph_weight}=[fill=white, draw=none, shape=rectangle, tikzit category=mbqc, inner sep=2pt, scale=.8]
\tikzstyle{graph_state}=[fill=yellow, draw=none, shape=rectangle, tikzit category=ZX, rounded corners=1.3mm, minimum height=1.7cm, minimum width=1.3cm, opacity=.7, text opacity=1]
\tikzstyle{box}=[fill=white, draw=black, shape=rectangle, inner sep=2.5pt, font={\scriptsize\boldmath}]
\tikzstyle{wirelable}=[font={\small}, fill=white, inner sep=1pt]
\tikzstyle{tightwirelable}=[font={\small}, fill=white, inner sep=0pt]

\tikzstyle{dash_edge}=[-, dashed]
\tikzstyle{hadamard_edge}=[-, dashed, dash pattern=on 2pt off 1.5pt, thick, draw=blue]
\tikzstyle{brace edge}=[-, tikzit draw=blue, decorate, decoration={brace,amplitude=1mm,raise=-1mm}]
\tikzstyle{thin}=[-, line width=0.5 pt]

%% file: introduction.tex
\section{Introduction}
\label{sec:introduction}

A helpful tool to reason about quantum computation is the
\emph{ZX-calculus}~\cite{coeckeInteractingQuantumObservables2011,coeckeInteractingQuantumObservables2008},
a graphical language which can represent any qubit computation.
It has been used, for example, in
measurement-based quantum computing~\cite{duncanRewritingMeasurementBasedQuantum2010, backensThereBackAgain2021, mcelvanneyCompleteFlowpreservingRewrite2022},
error-correcting codes~\cite{duncanVerifyingSteaneCode2014, garvieVerifyingSmallestInteresting2018, debeaudrapZXCalculusLanguage2020},
quantum circuit optimisation~\cite{debeaudrapFastEffectiveTechniques2020, duncanGraphtheoreticSimplificationQuantum2020, kissingerReducingNumberNonClifford2020},
classical simulation~\cite{kissingerClassicalSimulationQuantum2022, codsiClassicallySimulatingQuantum2023, laakkonenGraphicalSATAlgorithm2022},
quantum natural language processing~\cite{coeckeFoundationsNearTermQuantum2020, meichanetzidisQuantumNaturalLanguage2021},
quantum chemistry~\cite{shaikhHowSumExponentiate2022},
and quantum machine learning~\cite{wangDifferentiatingIntegratingZX2022, zhaoAnalyzingBarrenPlateau2021}.

All the above results use the \emph{qubit} ZX-calculus, but recent years have
seen a surge of interest in studying quantum computation using $d$-dimensional
systems, called \emph{qudits}.
Qudit-based quantum computation has been experimentally realised in a variety
of physical systems, such as
ion traps~\cite{ringbauerUniversalQuditQuantum2022, hrmoNativeQuditEntanglement2023},
photonic devices~\cite{chiProgrammableQuditbasedQuantum2022}, and
superconducting devices~\cite{blokQuantumInformationScrambling2021, yeCircuitQEDSinglestep2018,
    yurtalanImplementationWalshHadamardGate2020, hillRealizationArbitraryDoublycontrolled2021,
    gossHighfidelityQutritEntangling2022}.
On the theory side, there has been work in translating work on qubits to qudits
in quantum algorithms~\cite{wangQuditsHighDimensionalQuantum2020},
fault-tolerant quantum computing~\cite{gottesmanFaultTolerantQuantumComputation1999, campbellEnhancedFaultTolerantQuantum2014},
quantum communication~\cite{cozzolinoHighDimensionalQuantumCommunication2019},
and more~\cite{desilvaEfficientQuantumGate2021, gokhaleAsymptoticImprovementsQuantum2019,
    bocharovFactoringQutritsShor2017, nikolaevaDecomposingGeneralizedToffoli2022}.

This raises the question of how we can use the ZX-calculus to reason about
qudit systems. There exist several variations of the ZX-calculus that extend it
to higher-dimensional qudits. Many have focused on the specific case of qutrit
systems~\cite{wangQutritZXcalculusComplete2018, gongEquivalenceLocalComplementation2017,
wangQutritZXcalculusComplete2018,townsend-teagueSimplificationStrategiesQutrit2022}, with applications in quantum
computation~\cite{yehConstructingAllQutrit2022,
    vandeweteringPhaseGadgetCompilation2022}, and complexity theory~\cite{townsend-teagueSimplificationStrategiesQutrit2022}.
Recent papers have focused on the stabiliser fragment of odd prime dimensional qudits,
including Ref.~\cite{comfortAlgebraStabilizerCodes2023} that explores error correction and detection in this context, and also Ref.~\cite{boothCompleteZXcalculiStabiliser2022} mentioned below.
Some proposals capture all finite or infinite
dimensions~\cite{ranchinDepictingQuditQuantum2014, wangQufiniteZXcalculusUnified2022, poorCompletenessArbitraryFinite2023, defeliceLightmatterInteractionZXW2023},
but lack many of the nicer features of the qubit calculus.
Of particular importance to our paper is Ref.~\cite{boothCompleteZXcalculiStabiliser2022}, which
constructs a calculus for odd prime dimensions while retaining many of these
desirable properties and establishing completeness for the stabiliser fragment.
Despite these advancements, practical utilisation of the rewrites in
these calculi has received limited attention, leaving room for further
exploration and development.

To understand the usefulness of rewrite rules, we can take a look at the original qubit calculus.
In qubit ZX, we can distinguish between `standard' rules --- spider fusion, identity removal, state copying, bialgebra, and colour change --- and `harder' rules --- supplementarity, Euler angle colour permutation, and the rules dealing with the triangle generator.
The standard rules, with minor modifications, were those originally discovered~\cite{coeckeInteractingQuantumObservables2008}, and they are the most commonly used in practice.
For instance, all the rewrites used in the PyZX compiler~\cite{kissingerPyZXLargeScale2020} can be proved using just these standard rules~\cite{duncanGraphtheoreticSimplificationQuantum2020}.
These rules are sufficient to prove completeness for the \emph{stabiliser fragment} of the ZX-calculus~\cite{backensZXcalculusCompleteStabilizer2014}, while the harder rules were developed to prove completeness for larger fragments.
This suggests that carefully studying the qudit stabiliser fragment could be a fruitful avenue for developing useful qudit ZX rewrite rules.

Recall that the stabiliser fragment corresponds to Clifford computation, which is an efficiently simulable subset of quantum computation~\cite{gottesmanHeisenbergRepresentationQuantum1998} that forms the basis of many quantum protocols, such as error-correcting codes~\cite{kissingerPhasefreeZXDiagrams2022, khesinGraphicalQuantumCliffordencoder2023}, superdense coding~\cite{bennettCommunicationOneTwoparticle1992}, quantum teleportation~\cite{bennettTeleportingUnknownQuantum1993}, and quantum key distribution~\cite{bennettQuantumCryptographyPublic2014}.
Completeness of the qubit stabiliser fragment of ZX was proved in~\cite{backensZXcalculusCompleteStabilizer2014}, while for qutrits it was proved in~\cite{wangQutritZXcalculusComplete2018}.
Recently, completeness was proved for the stabiliser fragment for any odd-dimensional prime qudit dimension in~\cite{boothCompleteZXcalculiStabiliser2022}.
The proofs of all these results work essentially the same way: first, they show that any state diagram can be reduced to a Graph State with Local Cliffords (GSLC),
and then they show that any pair of GSLCs implementing the same state can be rewritten to a common reduced form.

In this paper, we take this last complete calculus for prime-dimensional
qudits~\cite{boothCompleteZXcalculiStabiliser2022} as a starting point, and
extend it in several ways:
\begin{enumerate}
    \item We simplify the rules to a smaller set that has a clearer relation to
    the original qubit stabiliser calculus, and for most of which we can prove the
    necessity.
    \item We incorporate a well-tempered axiomatisation for our calculus following
    the convention of~\cite{debeaudrapWelltemperedZXZH2021}, removing most of
    the scalars in our rewrite rules, and thus, simplifying our calculations.
    \item We introduce a new approach to handle scalars, formalising the
    often-used convention of writing scalar numbers alongside diagrams.
    \item We discover the qupit versions of the spider-removing \emph{local
    complementation} and \emph{pivoting} rules found in~\cite{
    duncanGraphtheoreticSimplificationQuantum2020} and generalised to qutrits
    in~\cite{townsend-teagueSimplificationStrategiesQutrit2022}.
    These rules serve as the foundation for optimisation and simulation
    strategies in the qubit setting~\cite{
        duncanGraphtheoreticSimplificationQuantum2020,
        kissingerReducingNumberNonClifford2020,
        debeaudrapFastEffectiveTechniques2020,kissingerPyZXLargeScale2020}.
    Our findings demonstrate that these strategies can be adapted to work for
    prime-dimensional qudits, thus extending their applicability beyond qubits.
    \item Using these rewrite rules, we simplify the original completeness proof
    of~\cite{boothCompleteZXcalculiStabiliser2022} by reducing the number of
    case distinctions required.%
    \footnote{
        In addition to being aesthetically and ergonomically preferable,
        reducing the number of case distinctions also makes the proof more
        easily verifiable. During the preparation of this manuscript, we
        identified and communicated several errors and omissions in~\cite{
        boothCompleteZXcalculiStabiliser2022}, which were subsequently fixed.
    }
    Specifically, we demonstrate that these rewrites reduce diagrams to a normal form that we call
    the \emph{affine with phases} (AP) form, which originally appeared in~\cite{
    dehaeneCliffordGroupStabilizer2003}.
    Then, given an AP-form diagram, we show how to reduce it further
    to a unique form, resulting in completeness.%
    \footnote{
        A similar normal form for qubits was independently found
        in~\cite{mcelvanneyCompleteFlowpreservingRewrite2022}.
        It is worth noting that our formulation was already employed for qubits in the
    Oxford Quantum Software course prior to the preprint~\cite{
        mcelvanneyCompleteFlowpreservingRewrite2022} appeared online.
    }
    \item Additionally, we demonstrate how to rewrite diagrams into a \emph{graph-state with
    local Cliffords} (GSLC) form, which yields a layered
    decomposition for Clifford unitaries similar to the one proposed for qubits
    in~\cite{duncanGraphtheoreticSimplificationQuantum2020}.
\end{enumerate}

Our findings highlight that qupit stabiliser diagrams share many familiar
properties with their qubit counterparts. Furthermore, many results regarding
optimisation and normal forms extend seamlessly to the odd prime-dimensional qudit
setting.

Finally, we have implemented many of these findings in \texttt{DiZX},
a new open-source Python library for qudit ZX-diagrammatic reasoning based on
\texttt{PyZX}~\cite{kissingerPyZXLargeScale2020}.%
\footnote{See \url{https://github.com/jvdwetering/dizx}.}

\paragraph{Related work}
Subsequent to submission, we were made aware of a related, parallel work,
Ref.~\cite{debeaudrapSimpleZXZH2023}, which also concerns well-tempered
axiomatisations for qudit ZX-calculi.

%% file: qupit-clifford-zx.tex
\section{The qupit Clifford ZX-calculus}
\label{sec:qupit-cliff}

In this section, we introduce the qudit stabiliser ZX-calculus for odd prime
dimensions.

We let $p$ denote an arbitrary odd prime, and $\Z_p = \Z/p\Z$ the ring of
integers modulo~$p$. Since $p$ is prime, $\Z_p$ is a field, implying that
every non-zero element in $\Z_p$ has a multiplicative inverse. We denote the
group of units (i.e.\@ invertible elements) as $\Z_p^* \coloneqq \Z_p \setminus \{0\}$.
We also define the Legendre symbol, for $x \in \Z_p^*$, as follows:
\begin{equation}
  \label{eq:legendre_characteristic}
  \left(\frac{x}{p}\right) =
  \begin{cases}
    1 \qif \exists y \in \Z_p^* \text{ s.t. } x = y^2; \\
    -1 \quad \text{otherwise};
  \end{cases}
\end{equation}

The Hilbert space of a qupit is $\mathcal{H} = \operatorname{span}\{\ket{m} \mid m \in \Z_p\} \cong \C^p$.
Letting $\omega \coloneqq e^{i\frac{2\pi}{p}}$ be a $p$-th primitive root
of unity, we can write down the following standard operators $Z$ and $X$, occasionally known as the \emph{clock}
and \emph{shift} operators:
$Z \ket{m} \coloneqq \omega^{m} \ket{m}$ and
$X \ket{m} \coloneqq \ket{m+1}$ for any $m \in \Z_p$. Notably, $ZX = \omega XZ$.

A \emph{Pauli operator} is defined as any operator of the form $\omega^k X^a Z^b$
for $k, a, b \in \Z_p$. We consider Pauli operator
\emph{trivial} if it is proportional to the identity. Each Pauli operator has a spectrum given by
$\{\omega^k \mid k \in \Z_p\}$, and we denote $\ket{k : Q}$ as the eigenvector
of a Pauli operator~$Q$ associated with the eigenvalue $\omega^k$.
It follows from the definition of $Z$ that we
can identify $\ket{k:Z} = \ket{k}$.

The collection of all Pauli operators is denoted $\mathscr{P}_1$ and called the
\emph{Pauli group}. For $n \in \N^*$, the \emph{generalised Pauli group}
$\mathscr{P}_n$ is defined as $\bigotimes_{k=1}^n \mathscr{P}_1$. Of particular
importance to us are the \emph{(generalised) Clifford groups}. These groups are
defined for each $n \in \N^*$ as the (unitary) normaliser of
$\mathscr{P}_n$. In other words, a unitary operator $C$ on
$\mathcal{H}^{\otimes n}$ belongs to the Clifford group if, for any
$P \in \mathscr{P}_n$, the conjugation $CPC^\dagger$ is also an element of
$\mathscr{P}_n$. While every Pauli operator is Clifford, there exist non-Pauli
Clifford operators.

In the case of prime qudit dimensions, the group of Clifford unitaries can be
generated by three gates: the \emph{Hadamard gate} defined as
$H \coloneqq \sum_{k \in \Z_p} \dyad{k:Z}{k:X}$, the $S$ gate defined as
$S \coloneqq \sum_{k \in \Z_p} \omega^{2^{-1} k(k-1)} \dyad{k:Z}{k:Z}$, and
the $CX$ gate defined as
$CX \coloneqq \sum_{j,k \in \Z_p} \dyad{j,j+k:Z}{j,k:Z}$~\cite{gottesmanFaultTolerantQuantumComputation1999}.
Note that in this context the Hadamard gate is sometimes also just called the \emph{Fourier transform}.

Stabiliser quantum mechanics is operationally described as a fragment of
quantum mechanics where the allowed operations include initialisations and
measurements in the eigenbases of Pauli operators, as well as unitary operations
from the generalised Clifford groups.

\subsection{Generators}

We define the symmetric monoidal category $\ZXp$ as having
objects $\N$ and morphisms generated by the following diagrams, for any $x,y
\in \Z_p$ and $s \in \C$:
\begin{align*}
  \input{generators/g-spider.tikz} &: m \to n &
  \input{generators/r-spider.tikz} &: m \to n &
  \input{generators/id.tikz} &: 1 \to 1 &
  \input{generators/hadamard.tikz} &: 1 \to 1 & \\
  \input{generators/cup.tikz} &: 0 \to 2 &
  \input{generators/cap.tikz} &: 2 \to 0 &
  \input{generators/braid.tikz} &: 2 \to 2 &
  \input{generators/scalar.tikz} &: 0 \to 0 &
\end{align*}
In addition to the \enquote{standard} generators of ZX, we have introduced a new generator represented by a light-grey bubble with a scalar written inside it, which we refer to as an \emph{explicit scalar}.
These explicit scalars offer a convenient way to streamline the often cumbersome reasoning related to scalars that is typically involved in many graphical completeness papers.
Note that the presence of the red X-spider as a generator is in principle unnecessary since the Z-spider surrounded by Hadamard boxes is equivalent to it.
However, our goal is not to provide a minimal set of generators, but rather a convenient one.

Diagrams in our framework can be composed in two ways:
sequentially, by connecting output wires to input wires,
or vertically, by \enquote{stacking} diagrams, corresponding to the tensor product operation
which is defined as $n \otimes m = n + m$ on objects.

\subsection{Interpretation}\label{subsec:interpretation}

The interpretation of a $\ZXp$-diagram is defined on objects as $\interp{m} \coloneqq \C^{p^m}$, and on the
generators as:
\begin{align*}
  \begin{aligned}
    \interp{\input{generators/g-spider.tikz}} &= p^{\frac{n+m-2}{4}}\sum_{k\in\Z_p} \omega^{2^{-1}(xk + yk^2)} \ket{k:Z}^{\otimes n}
    \bra{k:Z}^{\otimes m} &
    \interp{\input{generators/id.tikz}} &= \sum_{k\in\Z_p} \dyad{k:Z}{k:Z}  & \quad \\
    \interp{\input{generators/r-spider.tikz}} &= p^{\frac{n+m-2}{4}}\sum_{k\in\Z_p} \omega^{2^{-1}(xk + yk^2)} \ket{-k:X}^{\otimes n}
    \bra{k:X}^{\otimes m} &
    \interp{\input{generators/hadamard.tikz}} &= \sum_{k\in\Z_p} \dyad{k:Z}{k:X
    }  &
  \end{aligned} \\
  \begin{aligned}
    \interp{\!\!\!\input{generators/cup.tikz}\ } &= \sum_{k\in\Z_p} \ket{kk:Z} & \quad
    \interp{\ \input{generators/cap.tikz}\!\!\!} &= \sum_{k\in\Z_p} \bra{kk:Z} & \quad
    \interp{\ \input{generators/braid.tikz}\ } &= \sum_{k,\ell\in\Z_p} \dyad{k,
      \ell:Z}{\ell,k:Z}
  \end{aligned}
\end{align*}
and \(\interp{\input{generators/scalar.tikz}} = s\).

There are a couple of things we should remark about this interpretation.
First, the definition of the X-spider does not follow the standard convention.
It is defined in such a way that it maps X-eigenstates to their additive inverse (modulo $p$).
This definition is used in order to satisfy the property of \emph{flexsymmetry}~\cite{caretteWhenOnlyTopology2021,caretteWieldingZXcalculusFlexsymmetry2021},
which allows us to treat diagrams as undirected graphs.
Second, note that the interpretation of phases on the spiders has an additional $2^{-1}$ factor which
is necessary for the later stated \TextEuler and \TextGauss axioms to be sound.
This factor is considered modulo $p$, so for instance, for $p=5$ we have $2^{-1} \equiv 3$.
Finally, the spiders are defined with a global scalar factor of $p^{\frac{n+m-2}{4}}$ to follow the \emph{well-tempered normalisation} convention of~\cite{debeaudrapWelltemperedZXZH2021}.
This allows us to present the axioms later on with significantly fewer scalar factors floating around.

While the conventional qudit ZX-calculus represents spiders using a $(d - 1)$-dimensional vector~\cite{ranchinDepictingQuditQuantum2014},
we employ a different approach by leveraging a useful property of the Clifford group for prime-dimensional qudits:
the phases of its spiders are $p^m$-th roots of unity raised to polynomial functions with a maximum degree of $2$~\cite{cuiDiagonalGatesClifford2017}.
This property enables us to capture the essence of Clifford spiders using only two parameters: the coefficients of the linear and square terms.
As a result, we develop a more elegant and intuitive framework for reasoning about stabiliser maps, requiring only two parameters in any odd-prime dimension.
To establish a connection between our convention and the original qudit ZX-calculus,
we define a mapping where a spider with phase parameter $(x, y)$ corresponds to the spider described in~\cite{ranchinDepictingQuditQuantum2014} with parameter $\overrightarrow{\alpha} \coloneqq (\alpha_1, \cdots, \alpha_{d-1})$, where $\alpha_k = \omega^{2^{-1}(x k + y k^2)}$.

For any $a \in \Z_p$, the diagrams \input{new/spider-z-pauli.tikz} and \input{new/spider-x-pauli.tikz} correspond to the single qupit Pauli $Z^a$ and $X^a$ gates, respectively.
Similarly, the diagrams \input{new/spider-z-clifford.tikz} and \input{new/spider-x-clifford.tikz} correspond to Clifford unitaries for any $a, b \in \Z_p$.
As a result, we designate spiders with a phase $(a, 0)$ as \emph{Pauli spiders}, and spiders with a phase $(a, b)$ as \emph{Clifford spiders}.
Furthermore, spiders with a phase $(0, b)$ are referred to as \emph{purely-Clifford spiders}, while spiders with a phase $(a, z)$ where $z \neq 0$ are termed \emph{strictly-Clifford spiders}.
When the parameters of a spider are all zero, i.e.\@ $x = y = 0$, we call the spider \emph{phase-free} and we denote it without label as \input{new/phase_free_spider.tikz}, and similarly for the X-spider.
Lastly, we designate the phase-free X-spider \input{new/spider-antipode.tikz} as the \emph{antipode} since it implements the map $\ket{k:Z} \mapsto \ket{-k:Z}$.

Contrary to the qubit case, the qudit Hadamard gate is not self-inverse.
Instead, it follows the property that four successive applications of the
Hadamard gate results in the identity, that is, $H^4 = I$.
Therefore, the inverse of the Hadamard gate is given by $H^3$.
To maintain the clarity and simplicity of diagrams, we introduce the shorthand notation
$\input{equations/hadamard-inverse-simple.tikz}$ to represent the inverse of the Hadamard box.

\subsection{Axioms}\label{subsec:axioms}

\begin{figure}[tbh]
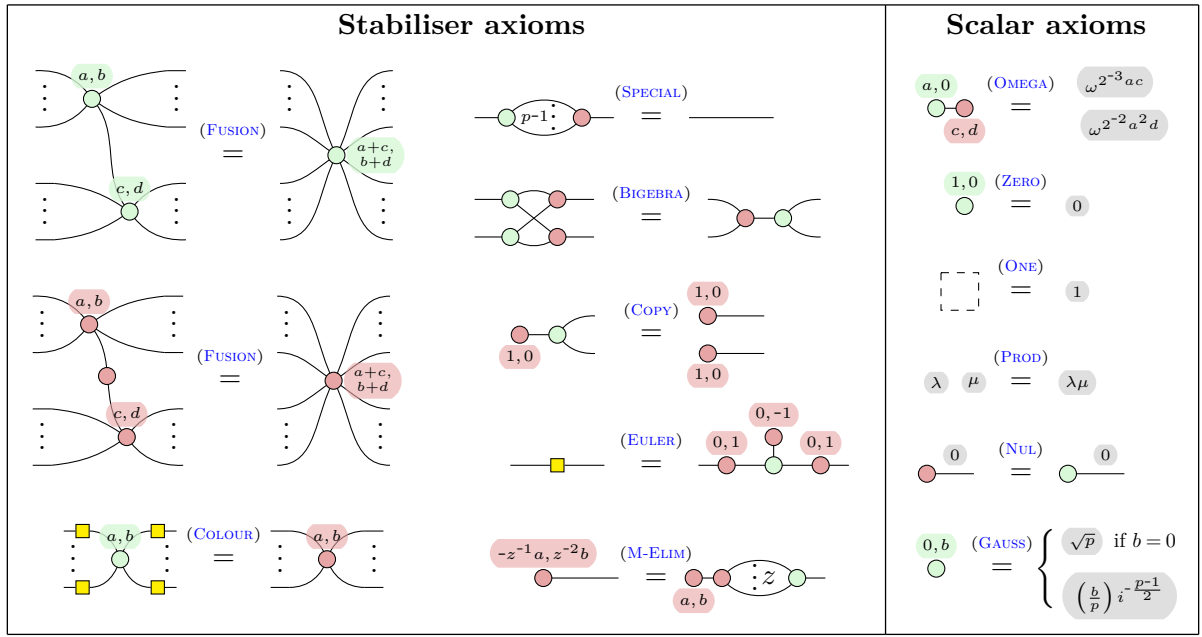

  \ctikzfig{axioms-new}
  \vspace{-0.3cm}
  \caption{The rewrite rules of the qudit stabiliser ZX-calculus for any odd
  prime dimension $p$. Here $a,b,c,d \in \Z_p$, \(z \in \Z_p^*\) and \(\lambda,\mu \in \C\).
    $\left(\frac{b}{p}\right)$ is the Legendre symbol, as defined in \cref{eq:legendre_characteristic}.
  The dotted square in \TextOne depicts the empty diagram.}
  \label{fig:axioms}
\end{figure}

We present the axioms of our calculus in \cref{fig:axioms}.
In addition to these concrete rules, our calculus also follows the structural rules of a compact-closed PROP\@.
This property implies that \enquote{only connectivity matters},
allowing us to treat our diagrams as undirected graphs while preserving their interpretation as linear maps.

These rewrite rules are essentially a simplified version of the complete set of
rewrite rules found in~\cite{boothCompleteZXcalculiStabiliser2022}.
We can show these rules are equivalent to those found in that paper, by deriving
the missing axioms.
\begin{proposition}\label{prop:missing-axioms}
  For any \( z \in \Z_p^* \) and \( a,c,d \in \Z_p \),
  $\ZXp$ proves the following axioms from~\cite{boothCompleteZXcalculiStabiliser2022}:
  \begin{equation*}
    \input{figures/equations/axioms-diff.tikz}
  \end{equation*}
\end{proposition}
\noindent Note that all the proofs in the paper can be found in the appendices.

We also change the presentation of scalars, but we can rely on the
reduction in~\cite{boothCompleteZXcalculiStabiliser2022} of the scalar fragment
to the elementary scalar fragment:
\begin{definition}
  \label{def:elementary-scalar}
  An \emph{elementary scalar} is a diagram \(A \in \ZXp[0,0]\) which is a
  (possibly empty) tensor product of diagrams from
  \(\{\input{figures/new/elementary_scalar_scalar.tikz},
  \input{figures/new/elementary_scalar_omega.tikz},
  \input{figures/new/elementary_scalar_sqrt_p.tikz},
  \input{figures/new/elementary_scalar_sqrt_p_inv.tikz},
  \input{figures/new/elementary_scalar_minus.tikz}
  \mid \lambda \in \C, s \in \Z_p\}\).
\end{definition}

\begin{restatable}{lemma}{elementaryScalarCompleteness}
  \label{lem:elementary_scalar_completeness}
  $\ZXp$ is complete for elementary scalars. Explicitly, if \(s: 0 \to 0\)
  is an elementary scalar, then \input{figures/scalar_completeness.tikz}.
\end{restatable}

With these results, we can see that every derivation of~\cite{boothCompleteZXcalculiStabiliser2022}
is also valid in our calculus, so that the rules of \cref{fig:axioms} are complete.
For this reason, we freely use the lemmas of~\cite{boothCompleteZXcalculiStabiliser2022} in the rest of this paper.

In deriving \textsc{Mult} and \textsc{Shear} in \cref{prop:missing-axioms}, as well as in the reduction to AP-form of
\cref{sec:ap-form}, we make extensive use of the following
\enquote{strictly-Clifford} state colour-change rules:
  \begin{restatable}{lemma}{lemStrictCliffordColour}
  \label{lem:state_colour_change}
  Strictly-Clifford states can all be represented both using Z- and X-spiders:
  for any \(a \in \Z_p\) and \(b \in \Z_p^*\),
  \begin{equation*}
    \input{figures/euler/state_colour_change.tikz}
  \end{equation*}
\end{restatable}
\noindent This lemma gives a qupit version of the well-known qubit ZX rule $\input{qubit-Y-rule.tikz}$.

On the way to proving this lemma, we also prove the qupit Clifford version of
\emph{supplementarity}, originally introduced for the qubit case in
Ref.~\cite{perdrixSupplementarityNecessaryQuantum2016}:
\begin{restatable}{lemma}{lemSupplementarity}
  \label{lem:fork_identity}
  For any \(b \in \Z_p^*\),
  \begin{equation*}
    \input{figures/euler/fork_identity.tikz}
  \end{equation*}
\end{restatable}
\noindent A generalisation of this rule is known to be necessary, but not
sufficient, for the completeness of the Clifford+T fragment in the qubit
case~\cite{perdrixSupplementarityNecessaryQuantum2016,
  jeandelZXCalculusCyclotomicSupplementarity2017}.

\subsection{A word on scalars}

Handling scalars in a graphical language is always a delicate issue.
Scalars are essential to guarantee the soundness of rewriting
rules but can sometimes be seen as a cumbersome bureaucracy that can be
omitted in practice and recovered through a quick normalisation check
at the end of a calculation. As a result, some textbooks prefer
to work up to non-zero scalars~\cite{coeckePicturingQuantumProcesses2017}, and
in~\cite{backensZXcalculusCompleteStabilizer2014}, a first proof of
completeness is presented without scalars, which are addressed in
a subsequent article~\cite{backensMakingStabilizerZXcalculus2015}.
There is no perfect solution to this situation.

In this paper, we adopt an intermediary approach that can be extended
to other graphical languages: the introduction of grey scalar
boxes. This approach bears resemblance to how the ZH-calculus
handles scalars~\cite{backensZHCompleteGraphical2019}, although in the
ZH-calculus, the scalar boxes are directly representable within the calculus itself,
requiring no extension as described here. Given any prop
$\textbf{P}$, the set of scalars $\textbf{P}[0,0]$ forms a commutative
monoid~\cite{heunenCategoriesQuantumTheory2019}.
We view $\textbf{P}[0,0]$ as a monoidal category with a single object, where
the $\otimes$ and $\circ$ operations are identified. We then consider the product category
$\textbf{P}[0,0] \times \textbf{P}$, which also forms a prop,
with arrows represented as pairs $(s,f)$, where $f: n \to m$ is an arrow of
$\textbf{P}$ and $s$ is a scalar.
Graphically, such a pair is depicted as a diagram representing $f$ together
with a floating grey scalar box containing $s$.
The principal equations governing the behaviour of scalar boxes are then
\TextOne and \TextProd.
In $\textbf{P}[0,0] \times \textbf{P}$, grey boxes and diagrams are treated independently.
To achieve the desired axiomatization of $\textbf{P}$, we need to quotient the
equational theory by the equation \input{figures/scalar_completeness.tikz} for all $s: 0 \to 0$.
This can be accomplished by introducing rules that guarantee the
desired result for a family of well-chosen elementary scalars.
Then it is enough to show that any diagram $0\to 0$ can be reduced to
elementary scalars as we do in Lemma~\ref{lem:elementary_scalar_completeness}.

%% file: figures/generators/g-spider.tikz
\begin{tikzpicture}
	\begin{pgfonlayer}{nodelayer}
		\node [style=gn, label={[gphase]below:$x,y$}] (0) at (0.075, 0) {};
		\node [style=none] (1) at (1.075, 0.5) {};
		\node [style=none] (2) at (1.075, -0.5) {};
		\node [style=none] (3) at (-0.925, 0.5) {};
		\node [style=none] (4) at (-0.925, -0.5) {};
		\node [style=none] (5) at (-1, 0) {$m$};
		\node [style=none] (6) at (-0.55, 0.2) {$\vdots$};
		\node [style=none] (7) at (1.05, 0) {$n$};
		\node [style=none] (8) at (0.675, 0.2) {$\vdots$};
	\end{pgfonlayer}
	\begin{pgfonlayer}{edgelayer}
		\draw [bend left] (0) to (1.center);
		\draw [bend right] (0) to (2.center);
		\draw [bend right] (0) to (3.center);
		\draw [bend left] (0) to (4.center);
	\end{pgfonlayer}
\end{tikzpicture}

%% file: figures/generators/r-spider.tikz
\begin{tikzpicture}
	\begin{pgfonlayer}{nodelayer}
		\node [style=rn, label={[rphase]below:$x,y$}] (9) at (0.075, 0) {};
		\node [style=none] (10) at (1.075, 0.5) {};
		\node [style=none] (11) at (1.075, -0.5) {};
		\node [style=none] (12) at (-0.925, 0.5) {};
		\node [style=none] (13) at (-0.925, -0.5) {};
		\node [style=none] (14) at (-1, 0) {$m$};
		\node [style=none] (15) at (-0.55, 0.2) {$\vdots$};
		\node [style=none] (16) at (1.05, 0) {$n$};
		\node [style=none] (17) at (0.675, 0.2) {$\vdots$};
	\end{pgfonlayer}
	\begin{pgfonlayer}{edgelayer}
		\draw [bend left] (9) to (10.center);
		\draw [bend right] (9) to (11.center);
		\draw [bend right] (9) to (12.center);
		\draw [bend left] (9) to (13.center);
	\end{pgfonlayer}
\end{tikzpicture}

%% file: figures/generators/id.tikz
\begin{tikzpicture}
	\begin{pgfonlayer}{nodelayer}
		\node [style=none] (0) at (-1, 0) {};
		\node [style=none] (1) at (1, 0) {};
	\end{pgfonlayer}
	\begin{pgfonlayer}{edgelayer}
		\draw (0.center) to (1.center);
	\end{pgfonlayer}
\end{tikzpicture}

%% file: figures/generators/hadamard.tikz
\begin{tikzpicture}
	\begin{pgfonlayer}{nodelayer}
		\node [style=had] (0) at (0, 0) {};
		\node [style=none] (1) at (-1, 0) {};
		\node [style=none] (2) at (1, 0) {};
	\end{pgfonlayer}
	\begin{pgfonlayer}{edgelayer}
		\draw (0) to (1.center);
		\draw (0) to (2.center);
	\end{pgfonlayer}
\end{tikzpicture}

%% file: figures/generators/cup.tikz
\begin{tikzpicture}
	\begin{pgfonlayer}{nodelayer}
		\node [style=none] (2) at (0, 0.5) {};
		\node [style=none] (3) at (0, -0.5) {};
		\node [style=none] (4) at (1, 0.5) {};
		\node [style=none] (5) at (1, -0.5) {};
		\node [style=none] (6) at (-1, 0.5) {};
		\node [style=none] (7) at (-1, -0.5) {};
	\end{pgfonlayer}
	\begin{pgfonlayer}{edgelayer}
		\draw [bend right=90, looseness=1.50] (2.center) to (3.center);
		\draw (4.center) to (2.center);
		\draw (5.center) to (3.center);
	\end{pgfonlayer}
\end{tikzpicture}

%% file: figures/generators/cap.tikz
\begin{tikzpicture}
	\begin{pgfonlayer}{nodelayer}
		\node [style=none] (0) at (0, 0.5) {};
		\node [style=none] (1) at (0, -0.5) {};
		\node [style=none] (2) at (-1, 0.5) {};
		\node [style=none] (3) at (-1, -0.5) {};
		\node [style=none] (4) at (1, 0.5) {};
		\node [style=none] (5) at (1, -0.5) {};
	\end{pgfonlayer}
	\begin{pgfonlayer}{edgelayer}
		\draw [bend left=90, looseness=1.50] (0.center) to (1.center);
		\draw (2.center) to (0.center);
		\draw (3.center) to (1.center);
	\end{pgfonlayer}
\end{tikzpicture}

%% file: figures/generators/braid.tikz
\begin{tikzpicture}
	\begin{pgfonlayer}{nodelayer}
		\node [style=none] (0) at (-1, 0.5) {};
		\node [style=none] (3) at (1, -0.5) {};
		\node [style=none] (4) at (-1, -0.5) {};
		\node [style=none] (5) at (1, 0.5) {};
	\end{pgfonlayer}
	\begin{pgfonlayer}{edgelayer}
		\draw [in=-180, out=0, looseness=1.75] (0.center) to (3.center);
		\draw [in=180, out=0, looseness=1.50] (4.center) to (5.center);
	\end{pgfonlayer}
\end{tikzpicture}

%% file: figures/generators/scalar.tikz
\begin{tikzpicture}
	\begin{pgfonlayer}{nodelayer}
		\node [style=scalar] (0) at (0, 0) {$s$};
	\end{pgfonlayer}
\end{tikzpicture}

%% file: figures/new/spider-z-pauli.tikz
\begin{tikzpicture}
	\begin{pgfonlayer}{nodelayer}
		\node [style=none] (0) at (-1, 0) {};
		\node [style=none] (1) at (1, 0) {};
		\node [style=gn, label={[gphase]below:$a,0$}] (2) at (0, 0) {};
	\end{pgfonlayer}
	\begin{pgfonlayer}{edgelayer}
		\draw (0.center) to (2);
		\draw (2) to (1.center);
	\end{pgfonlayer}
\end{tikzpicture}

%% file: figures/new/spider-x-pauli.tikz
\begin{tikzpicture}
	\begin{pgfonlayer}{nodelayer}
		\node [style=rn, label={[rphase]below:$a,0$}] (0) at (0, 0) {};
		\node [style=none] (1) at (-1, 0) {};
		\node [style=none] (2) at (2, 0) {};
		\node [style=rn] (3) at (1, 0) {};
	\end{pgfonlayer}
	\begin{pgfonlayer}{edgelayer}
		\draw (2.center) to (1.center);
	\end{pgfonlayer}
\end{tikzpicture}

%% file: figures/new/spider-z-clifford.tikz
\begin{tikzpicture}
	\begin{pgfonlayer}{nodelayer}
		\node [style=none] (0) at (-1, 0) {};
		\node [style=none] (1) at (1, 0) {};
		\node [style=gn, label={[gphase]below:$a,b$}] (2) at (0, 0) {};
	\end{pgfonlayer}
	\begin{pgfonlayer}{edgelayer}
		\draw (0.center) to (2);
		\draw (2) to (1.center);
	\end{pgfonlayer}
\end{tikzpicture}

%% file: figures/new/spider-x-clifford.tikz
\begin{tikzpicture}
	\begin{pgfonlayer}{nodelayer}
		\node [style=rn, label={[rphase]below:$a,b$}] (0) at (0, 0) {};
		\node [style=none] (1) at (-1, 0) {};
		\node [style=none] (2) at (2, 0) {};
		\node [style=rn] (3) at (1, 0) {};
	\end{pgfonlayer}
	\begin{pgfonlayer}{edgelayer}
		\draw (2.center) to (1.center);
	\end{pgfonlayer}
\end{tikzpicture}

%% file: figures/new/phase_free_spider.tikz
\begin{tikzpicture}
	\begin{pgfonlayer}{nodelayer}
		\node [style=none] (0) at (-1, 0) {};
		\node [style=none] (1) at (1, 0) {};
		\node [style=gn] (2) at (0, 0) {};
	\end{pgfonlayer}
	\begin{pgfonlayer}{edgelayer}
		\draw (1.center) to (0.center);
	\end{pgfonlayer}
\end{tikzpicture}

%% file: figures/new/spider-antipode.tikz
\begin{tikzpicture}
	\begin{pgfonlayer}{nodelayer}
		\node [style=none] (0) at (-1, 0) {};
		\node [style=none] (1) at (1, 0) {};
		\node [style=rn] (2) at (0, 0) {};
	\end{pgfonlayer}
	\begin{pgfonlayer}{edgelayer}
		\draw (1.center) to (0.center);
	\end{pgfonlayer}
\end{tikzpicture}

%% file: figures/new/elementary_scalar_scalar.tikz
\begin{tikzpicture}
	\begin{pgfonlayer}{nodelayer}
		\node [style=scalar] (0) at (0, 0) {$\lambda$};
	\end{pgfonlayer}
\end{tikzpicture}

%% file: figures/new/elementary_scalar_omega.tikz
\begin{tikzpicture}
	\begin{pgfonlayer}{nodelayer}
		\node [style=gn, label={[gphase]right:$s,0$}] (0) at (-0.375, 0) {};
		\node [style=rn, label={[rphase]left:$1,0$}] (1) at (0.375, 0) {};
	\end{pgfonlayer}
	\begin{pgfonlayer}{edgelayer}
		\draw (1) to (0);
	\end{pgfonlayer}
\end{tikzpicture}

%% file: figures/new/elementary_scalar_sqrt_p.tikz
\begin{tikzpicture}
	\begin{pgfonlayer}{nodelayer}
		\node [style=gn] (0) at (-0.375, 0) {};
		\node [style=rn] (1) at (0.375, 0) {};
	\end{pgfonlayer}
	\begin{pgfonlayer}{edgelayer}
		\draw (0) to (1);
	\end{pgfonlayer}
\end{tikzpicture}

%% file: figures/new/elementary_scalar_sqrt_p_inv.tikz
\begin{tikzpicture}
	\begin{pgfonlayer}{nodelayer}
		\node [style=gn] (0) at (-0.75, 0) {};
		\node [style=rn] (1) at (0.75, 0) {};
		\node [style=rn] (2) at (0, 0) {};
	\end{pgfonlayer}
	\begin{pgfonlayer}{edgelayer}
		\draw [bend right=60] (0) to (1);
		\draw (2) to (0);
		\draw (2) to (1);
		\draw [bend left=60] (0) to (1);
	\end{pgfonlayer}
\end{tikzpicture}

%% file: figures/new/elementary_scalar_minus.tikz
\begin{tikzpicture}
	\begin{pgfonlayer}{nodelayer}
		\node [style=gn, label={[gphase]left:$0,1$}] (0) at (0, 0) {};
	\end{pgfonlayer}
\end{tikzpicture}

%% file: figures/scalar_completeness.tikz
\begin{tikzpicture}
	\begin{pgfonlayer}{nodelayer}
		\node [style=box] (0) at (-1, 0) {s};
		\node [style=scalar] (1) at (1.25, 0) {$\interp{s}$};
		\node [style=none] (2) at (0, 0) {$=$};
	\end{pgfonlayer}
\end{tikzpicture}

%% file: figures/qubit-Y-rule.tikz
\begin{tikzpicture}
	\begin{pgfonlayer}{nodelayer}
		\node [style=rn] (0) at (1.25, 0) {};
		\node [style=none] (1) at (2.25, 0) {};
		\node [style=none] (2) at (0, 0) {$\propto$};
		\node [style=gn] (3) at (-1.95, 0) {};
		\node [style=none] (4) at (-0.95, 0) {};
		\node [style=gphase] (5) at (-1.95, 0.625) {$\pm\frac\pi2$};
		\node [style=rphase] (10) at (1.25, 0.675) {$\mp\frac\pi2$};
	\end{pgfonlayer}
	\begin{pgfonlayer}{edgelayer}
		\draw (0) to (1.center);
		\draw (3) to (4.center);
	\end{pgfonlayer}
\end{tikzpicture}

%% file: ap-form.tex
\section{Normal forms}
\label{sec:ap-form}

In this section, we show that we can simplify stabiliser diagrams
into two distinct normal forms:
the \emph{affine with phases} (AP) form and
the \emph{graph state with local Cliffords} (GSLC) form.
The AP form can be efficiently transformed into a unique reduced form,
offering an alternative proof of completeness.
On the other hand, the GSLC form is particularly useful for rewriting and
decomposing stabiliser unitaries.

\subsection{Graph simplifications}\label{subsec:graph-simplifications}

Before reducing the diagrams to our normal forms,
we first need to simplify them into a \emph{graph-like} form.
In this form, the diagrams consist only of Z-spiders and \emph{H-edges}.
To define the qupit graph-like diagrams,
we first define \emph{H-boxes} as:
\begin{equation*}
  \label{eq:hbox_def}
  \input{equations/weighted_hadamard_multiedge.tikz}
\end{equation*}
where $x \in \Z_p$ is the \emph{weight} of the H-box.
Unlike the \emph{multipliers} in~\cite{boothCompleteZXcalculiStabiliser2022},
H-boxes are undirected, thus, we can treat diagrams that contain only
generators and H-boxes as undirected (weighted) graphs.

\begin{restatable}{proposition}{prophbox}
  \label{prop:hbox}
  $\ZXp$ proves the following equations:
  \begin{equation*}
    \input{equations/weighted_hadamard.tikz}
  \end{equation*}
\end{restatable}

Since edges that contain H-boxes are central to the subsequent proofs, we
define \emph{H-edges}, similarly to the qubit case, as a blue dashed line
with the corresponding weight on top:
\begin{equation}
  \label{eq:xhad-edge-def}
  \input{new/hadamard-edge-def.tikz}
\end{equation}

\begin{definition}
  A ZX-diagram is \emph{graph-like} when:
  \begin{enumerate}
    \item All spiders are Z-spiders.
    \item Z-spiders are only connected via H-edges.
    \item There are no self-loops.
    \item Every input or output is connected to a Z-spider.
    \item Every Z-spider is connected to at most one input or output.
  \end{enumerate}
\end{definition}

Using standard techniques~\cite{duncanGraphtheoreticSimplificationQuantum2020},
it is evident that any ZX-diagram can be transformed into a graph-like form.
This transformation involves several steps:
performing a colour change on all X-spiders,
fusing all Z-spiders, removing self-loops,
and introducing identity elements to ensure that each input and output is
correctly connected to a Z-spider.
Once in graph-like form, the diagram can be represented as an open, weighted graph,
where the edge weights are elements of $\mathbb{Z}_p$ and each vertex is
labelled by a phase $(a, b) \in \mathbb{Z}_p^2$.

Now that we have a graph-like diagram, we can differentiate between \emph{boundary}
spiders, those directly connected to an input or output, and \emph{interior}
spiders, those that are only connected to other spiders.
Subsequently, we demonstrate that many of the internal spiders can be removed
from a diagram using similar techniques to the qubit case~\cite{
  duncanGraphtheoreticSimplificationQuantum2020}.

The local complementation simplification enables the removal of a
strictly-Clifford interior spider by introducing phases and wires to the
spiders it is connected to.
This technique is analogous to the qubit version described
in~\cite{duncanGraphtheoreticSimplificationQuantum2020}.

\begin{restatable}[Local complementation simplification]{lemma}{lc}
  \label{lem:lc}
  For any $z \in \Z^*_p$ and for all $a, \alpha_i, \beta_i, e_i, w_{i,j} \in \Z_p$ where
  $i, j \in \{1, \ldots k\}$ such that $i < j$ we have:
  \begin{equation*}
    \input{new/local_compl.tikz}
  \end{equation*}
  Here
  $\gamma_i = \alpha_i - e_i a z^{\minu 1}$,
  $\delta_i = \beta_i - z^{\minu 1} e_i^2$,
  and $g_{i,j} = w_{i j} - z^{\minu 1} e_i e_j$.
\end{restatable}

We also have an analogue of the pivot rewrite rule.
This rule enables us to eliminate connected interior Pauli spiders by
introducing additional phases and connections to the spiders they are connected to.

First, we prove a simplified version of pivoting:
\begin{restatable}{lemma}{pivotpartial}
  \label{lem:pivot-partial}
  The following version of pivoting is derivable in $\ZXp$:
  \begin{equation*}
    \input{new/partial_pivot.tikz}
  \end{equation*}
  Here $\epsilon \in \Z^*_p$ and all the other variables are allowed
  arbitrary values.
\end{restatable}
Then the general version can be derived from that:
\begin{restatable}[Pivoting simplification]{lemma}{pivot}
  \label{lem:pivot}
  General pivoting is derivable in $\ZXp$:
  \begin{equation*}
    \input{new/pivot.tikz}
  \end{equation*}
  Here again $\epsilon \in \Z^*_p$ with every other variable on the left-hand side
  allowed arbitrary values.
  On the right-hand side
  $\gamma_i = \alpha_i - \epsilon^{\minu 1} (a f_i + b e_i)$,
  $\delta_i = \beta_i - 2 \epsilon^{\minu 1} e_i f_i$,
  and $g_{i,j} = - \epsilon^{\minu 1} (e_i f_j + e_j f_i)$.
\end{restatable}

\subsection{AP-form}\label{subsec:ap-form}

The above results suggest that through the application of local complementation
and pivoting, it is possible to transform any state diagram (a diagram without
inputs) into a graph-like diagram where only Pauli spiders remain internal
spiders, and they are exclusively connected to boundary spiders.
This is achieved through a two-step process.
Firstly, any internal spider that is Clifford is eliminated through local complementation.
This ensures that only Pauli spiders remain internal.
Secondly, given that the diagram contains only Pauli internal spiders,
any connected pair of internal spiders can be removed using pivoting.
We give a name to this type of diagram:
\begin{definition}
  We say that a graph-like diagram is in \emph{Affine with Phases form}
  (AP-form) when:
  \begin{itemize}
    \item There are no inputs;
    \item The internal spiders are Pauli spiders;
    \item Internal spiders are only connected to boundary spiders.
  \end{itemize}
\end{definition}

We refer to this class of diagrams as \enquote{Affine with Phases} because they
correspond to states described by an affine subspace of $Z$ basis states,
with an additional phase function applied to the output.
This characterisation is supported by the following lemma:

\begin{restatable}{lemma}{apform}
  A general non-zero $n$-qupit diagram in AP-form is described by the diagram:
  \begin{equation}
    \label{eq:ap-form}
    \input{figures/new/ap_form.tikz}
  \end{equation}
  where $a_l, \alpha_i, \beta_i, e_{h,i}, f_{i,j} \in \Z_p$ with
  $l \in \{1, \ldots, k\}$ and $i,j \in \{1, \ldots, n\}$ such that $i < j$.
  The interpretation of this diagram is (up to some non-zero scalar) equal to a state
  \begin{equation}
    \label{eq:ap-form-poly}
    \sum_{E \vec{x} = \vec{a}}
    \omega^{\phi(\vec x)} \ket{\vec x}
  \end{equation}
  where $E$ is the weighted bipartite adjacency matrix of the internal and boundary spiders,
  $\vec a$ describes the Pauli phases of the internal spiders, and $\phi$ is a
  phase function that describes the connectivity and phases of the boundary spiders:
  \begin{equation*}
    \label{eq:ap-form-matrices}
    E =
    \begin{bmatrix}
      e_{1,1} & \cdots & e_{1,n} \\
      e_{2,1} & \cdots & e_{2,n} \\
      \vdots  &        & \vdots     \\
      e_{k,1} & \cdots & e_{k,n} \\
    \end{bmatrix}
    \ ,\qquad
    \vec a =
    \begin{bmatrix}
      a_1    \\
      \vdots \\
      a_k
    \end{bmatrix}
    \ ,\qquad
    \phi(\vec x) =
    \sum_{\substack{i, j \in \{ 1, \ldots, n \} \\ i < j}}
    2^{\minu 3} x_i \alpha_i + 2^{\minu 2} x_i^2 \beta_i - 2^{\minu 3} f_{i,j} x_i x_j
  \end{equation*}
\end{restatable}

Notably, states described by AP-form diagrams correspond to the stabiliser
normal forms described in Ref.\@~\cite{vandennestClassicalSimulationQuantum2010}.


With AP-form diagrams, we can prove a qupit version of the Gottesman-Knill
theorem, which states that we can efficiently sample from the probability distribution
of a stabiliser computation. Let us consider an AP-form diagram represented
by $(E, \vec{b}, \phi)$. When we measure this state in the computational basis,
we observe that the phase function $\phi$ has no impact on the measurement
outcomes, allowing us to disregard it. Hence, we can describe the state as
$N\sum_{E\vec{x} = \vec{a}} \ket{\vec{x}}$, where $N$ is a normalisation
constant. This state represents a uniform superposition of the states $\ket{\vec{x}}$
that satisfy the equation $E\vec{x} = \vec{a}$.

To sample from such states, we need to generate solutions to this equation
uniformly at random. Efficiently achieving this involves finding any solution
$E\vec{x}' = \vec{a}$ and then obtaining a basis $\vec{v}_1, \ldots, \vec{v}_\ell$
for the linear space $\{E\vec{x} = \vec{0}\}$. We can then return
$\vec{x}' + \sum_i^{\ell} b_i \vec{v}_i$, where the $b_i \in \mathbb{Z}_p$ are
chosen uniformly at random.

AP-form diagrams also enable us to provide an alternative, more direct proof of
the completeness of $\ZXp$ through reduction to a unique normal form.
In the context of graphical calculi, completeness means that the rewrite rules
of the calculus can prove any true equation. In other words, if
$\interp{A} = \interp{B}$, then it is possible to rewrite diagram $A$ into
diagram $B$.

\begin{definition}
We say that a diagram in AP-form defined by $(E, \vec a, \phi)$ is in
\emph{reduced AP-form} if it is either zero, or it is non-zero and satisfies
the following conditions:
\begin{itemize}
  \item $E$ is in reduced row echelon form (RREF), i.e., it is fully reduced using
  Gaussian elimination.
  \item $E$ contains no fully zero rows.
  \item $\phi$ only contains free variables from the equation system of $E$,
  i.e.\@, variables that do not correspond to \emph{pivot} columns in $E$.
\end{itemize}
\end{definition}

\begin{restatable}{lemma}{apunique}
  \label{lem:ap-unique}
  For any non-zero state $\ket \psi$, there is at most one triple
  $(E, \vec a, \phi)$ satisfying the conditions of reduced AP-form such that:
  \begin{equation*}
    \ket \psi
    \approx
    \sum_{E \vec{x} = \vec{a}}
    \omega^{\phi(\vec x)} \ket{\vec x}
  \end{equation*}
\end{restatable}
\noindent Therefore, a diagram in reduced AP-form is unique.

Now, our objective is to demonstrate that we can rewrite a ZX-diagram in
AP-form in a manner that transforms its biadjacency matrix $E$ into RREF\@.
Additionally, we need to show that we can modify the diagram so that the
corresponding phase function $\phi$ only includes free variables from the
equation system $E \vec{x} = \vec{a}$.
Put simply, we need to prove that we can perform primitive row operations on a
ZX-diagram in AP-form as well as eliminate any phase or Hadamard edge from a
pivot spider.

\begin{restatable}{lemma}{rowadd}
  \label{lem:row-add}
  We can perform primitive row operations on a ZX-diagram in AP-form, i.e., we can ``add'' one inner spider to another. For any $k, a, b, e_i, f_j \in \Z_p$ where $i \in {1, \ldots, n}$ and $j \in {1, \ldots, m}$:
  \begin{equation*}
  \input{figures/new/row_add.tikz}
  \end{equation*}
\end{restatable}

Using this result, we can apply primitive row operations to $E$ in AP-form
diagram and hence reduce it to RREF\@.
Through diagrammatic rewrites, we can show that when $E$ is in RREF,
we can eliminate all the phases and H-edges associated with the non-free
variables of $E$.

\begin{restatable}{lemma}{removephasehad}
  \label{lem:remove_phase_had}
  If an AP-form diagram has its biadjacency matrix $E$ in RREF, we can rewrite
  the diagram so that the boundary spiders corresponding to non-free variables
  of $E$ have zero phases, and there are no H-edges connecting them to other
  boundary spiders.
\end{restatable}

\begin{restatable}{lemma}{reducedap}
  \label{lem:reduced-ap}
  Any diagram in $\ZXp$ can be converted into one in reduced AP-form.
\end{restatable}

The completeness result follows immediately from the above lemma.

\begin{restatable}[Completeness]{theorem}{completeness}
  For any pair of ZX-diagrams $A, B \in \ZXp$, if $\interp{A} = \interp{B}$,
  we can provide a sequence of rewrites that transforms $A$ into $B$.
\end{restatable}

\subsection{GSLC form}

The AP-form is advantageous as it can be directly transformed into a unique normal form,
and allows for straightforward classical sampling.
However, it may be less suitable for other applications.
For instance, when applying the algorithm described above to a diagram originating from a Clifford unitary,
it becomes challenging to establish a clear relationship between the resulting simplified diagram and a corresponding quantum circuit.

In this section, we introduce the qupit version of the well-known qubit GSLC-form diagrams.
\begin{definition}
  We say a diagram is in \emph{GSLC form} (Graph State with Local Cliffords)
  when it is graph-like, up to Hadamards on input and output wires, and it
  has no internal spiders.
\end{definition}

The algorithm for reducing a diagram to AP-form may still yield diagrams
with internal spiders, specifically Pauli spiders connected to boundaries.
However, we can eliminate these internal spiders by using a \emph{boundary pivot}.

\begin{restatable}{lemma}{boundarypivot}
  \label{lem:boundary-pivot}
  The following boundary pivot rule is derivable in $\ZXeq$:
  \ctikzfig{new/partial-pivot-boundary}
  Here $g_{ij} \coloneqq -\epsilon^{-1}e_if_j$ and $h_i \coloneqq -\epsilon^{-1}e_i$.
  This rule holds for all choices of phases as long as $\epsilon \neq 0$.
\end{restatable}

To observe how this rewrite aids in eliminating internal spiders, consider that the
spider with a phase of $(b,c)$ now becomes an internal spider connected to an internal
Pauli spider. Consequently, if $c=0$, we can eliminate the pair using standard
pivoting. On the other hand, if $c\neq 0$, we can employ a local complementation to remove
the $(b,c)$ spider. This alteration modifies the phase of its sole neighbour,
subsequently enabling its removal through another local complementation.

Lemma~\ref{lem:boundary-pivot} can be straightforwardly modified, similar to
Lemma~\ref{lem:pivot}, to accommodate arbitrary connectivity between the
internal spider and the boundary. By incorporating additional spider unfusions,
we can extend the application of Lemma~\ref{lem:boundary-pivot} to boundary spiders
that are connected to multiple inputs or outputs. It is worth noting that when
applying \cref{lem:boundary-pivot} multiple times to the same boundary,
different powers of the Hadamard gate may appear on the input or output wire.
For instance, applying it twice yields $(H^3)^2 = H^2$, and another iteration
reverts back to $H$.

Hence, we can observe that it is indeed possible to eliminate all internal spiders from a diagram,
allowing for an efficient reduction of diagrams to GSLC form. This is particularly
significant for diagrams derived from unitaries, as we can then
rewrite them in the following manner:
\begin{equation*}
  \input{GSLC-extract-1.tikz}
\end{equation*}
Here, the boxes labelled with $H?$ represent a possible power of a Hadamard gate acting on the
qupit. By applying spider unfusion and colour change operations, we observe that the diagram
can be decomposed into several layers consisting of Hadamard gates, Z phase
gates, CZ gates, and a middle portion represented by a weighted biadjacency matrix $A$.
This part of the circuit implements a map of the form
$\ket{\vec x} \mapsto \ket{A \vec x}$, where $\vec x \in \mathbb{Z}_p^n$ and
$A$ is an $n\times n$ matrix over $\mathbb{Z}_p$.
Since we assume the entire map to be unitary, $A$ must also be invertible.
Consequently, such a `linear' qupit map can always be implemented through a series of CX
gates, transforming $\ket{x,y}$ to $\ket{x,x+y}$ (the decomposition is achieved via
standard Gaussian elimination over $\mathbb{Z}_p$).
Thus, we arrive at the following result.
\begin{theorem}
  Any odd-prime-dimensional qudit Clifford unitary can be efficiently
  decomposed into a quantum circuit consisting of the following layers:
  \begin{quote}
    H---Z---S---CZ---CX---H---CZ---Z---S---H
  \end{quote}
\end{theorem}
To the best of our knowledge, such a Clifford normal form for qudits has not been
described before in the existing literature.
It is worth noting, though, that this result bears a striking resemblance to the qubit normal form
for Clifford circuits outlined in~\cite{duncanGraphtheoreticSimplificationQuantum2020}.

%% file: figures/GSLC-extract-1.tikz
\begin{tikzpicture}
	\begin{pgfonlayer}{nodelayer}
		\node [style=box] (0) at (-10, 1.75) {H?};
		\node [style=none] (1) at (-11.75, 1.75) {};
		\node [style=none] (2) at (-11.75, 0.25) {};
		\node [style=none] (3) at (-11.75, -1.75) {};
		\node [style=box] (4) at (-10, 0.25) {H?};
		\node [style=box] (5) at (-10, -1.75) {H?};
		\node [style=none, rotate=90] (9) at (-10, -0.75) {...};
		\node [style=none] (10) at (-1.25, -1.75) {};
		\node [style=box] (11) at (-3, 1.75) {H?};
		\node [style=none] (12) at (-1.25, 0.25) {};
		\node [style=box] (13) at (-3, 0.25) {H?};
		\node [style=none] (14) at (-1.25, 1.75) {};
		\node [style=box] (18) at (-3, -1.75) {H?};
		\node [style=none] (20) at (-7.25, 1.75) {};
		\node [style=none] (21) at (-7.25, 0.5) {};
		\node [style=none] (22) at (-7.25, -0.5) {};
		\node [style=none] (23) at (-7.25, -1.5) {};
		\node [style=none] (24) at (-7.25, -1.75) {};
		\node [style=none] (25) at (-7.25, 0) {};
		\node [style=none] (26) at (-7.25, 1.25) {};
		\node [style=none] (27) at (-5.75, -1.75) {};
		\node [style=none] (28) at (-5.75, 1.75) {};
		\node [style=none] (29) at (-5.75, -1.75) {};
		\node [style=none] (30) at (-5.75, 1.25) {};
		\node [style=none] (31) at (-5.75, 0.75) {};
		\node [style=none] (32) at (-5.75, -0.5) {};
		\node [style=none] (33) at (-5.75, 0) {};
		\node [style=none] (34) at (-5.75, -1.25) {};
		\node [style=none, rotate=90] (35) at (-6.5, 0) {\Huge$\cdots$};
		\node [style=none] (36) at (0, 0) {$=$};
		\node [style=none] (37) at (10, 0.75) {};
		\node [style=none] (38) at (10, 1.75) {};
		\node [style=none] (39) at (10, -1.25) {};
		\node [style=none] (42) at (1.25, 1.75) {};
		\node [style=rn] (43) at (11, -1.75) {};
		\node [style=none] (45) at (8.5, -1) {};
		\node [style=none] (46) at (1.25, 0.25) {};
		\node [style=none] (48) at (17.25, -1.75) {};
		\node [style=none] (50) at (10, 0) {};
		\node [style=none] (51) at (8.5, -0.5) {};
		\node [style=none] (52) at (8.5, 0.5) {};
		\node [style=none] (54) at (8.5, 1.75) {};
		\node [style=none] (55) at (10, -0.5) {};
		\node [style=gn] (56) at (7.5, -1.75) {};
		\node [style=none] (57) at (8.5, 1.25) {};
		\node [style=gn] (58) at (7.5, 1.75) {};
		\node [style=none] (59) at (8.5, -1.75) {};
		\node [style=gn] (60) at (7.5, 0.25) {};
		\node [style=none] (61) at (10, 1.25) {};
		\node [style=none] (62) at (17.25, 0.25) {};
		\node [style=rn] (64) at (11, 1.75) {};
		\node [style=rn] (65) at (11, 0.25) {};
		\node [style=none] (66) at (17.25, 1.75) {};
		\node [style=none] (67) at (1.25, -1.75) {};
		\node [style=none] (68) at (8.5, 0) {};
		\node [style=none] (69) at (10, -1.25) {};
		\node [style=none, rotate=90] (71) at (9.25, 0) {\Huge$\cdots$};
		\node [style=none] (72) at (10, -1.75) {};
		\node [style=gn] (73) at (6.5, 1.75) {};
		\node [style=gn] (74) at (6.5, 0.25) {};
		\node [style=gn] (75) at (5.5, 1.75) {};
		\node [style=gn] (76) at (5.5, -1.75) {};
		\node [style=gn] (77) at (13.25, 0.25) {};
		\node [style=gn] (78) at (13.25, -1.75) {};
		\node [style=had] (79) at (5.5, -1) {$x$};
		\node [style=had] (80) at (6.5, 1) {$y$};
		\node [style=had] (81) at (13.25, -0.75) {$z$};
		\node [style=had] (82) at (12, 1.75) {};
		\node [style=had] (83) at (12, 0.25) {};
		\node [style=had] (84) at (12, -1.75) {};
		\node [style=none] (85) at (11.5, -2.375) {};
		\node [style=none] (86) at (7, -2.375) {};
		\node [style=none] (87) at (9.25, -3) {$A$};
		\node [style=box] (88) at (2.5, 1.75) {H?};
		\node [style=box] (89) at (2.5, 0.25) {H?};
		\node [style=box] (90) at (2.5, -1.75) {H?};
		\node [style=box] (91) at (15.75, 1.75) {H?};
		\node [style=box] (92) at (15.75, 0.25) {H?};
		\node [style=box] (93) at (15.75, -1.75) {H?};
		\node [style=none, rotate=90] (94) at (-3, -0.75) {...};
		\node [style=none, rotate=90] (95) at (2.5, -0.75) {...};
		\node [style=none, rotate=90] (96) at (15.75, -0.75) {...};
		\node [style=wirelable] (97) at (-8.75, 0.75) {$x$};
		\node [style=wirelable] (98) at (-8.25, 1) {$y$};
		\node [style=wirelable] (99) at (-4.75, -0.75) {$z$};
		\node [style=gn, label={[gphase]below:$a_1,b_1$}] (100) at (-8.25, 1.75) {};
		\node [style=gn, label={[gphase]above:$a_2,b_2$}] (101) at (-8.25, 0.25) {};
		\node [style=gn, label={[gphase]above:$a_n,b_n$}] (102) at (-8.25, -1.75) {};
		\node [style=gn, label={[gphase]below:$c_1,d_1$}] (103) at (-4.75, 1.75) {};
		\node [style=gn, label={[gphase]below:$c_2,d_2$}] (104) at (-4.75, 0.25) {};
		\node [style=gn, label={[gphase]above:$c_n,d_n$}] (105) at (-4.75, -1.75) {};
		\node [style=gn, label={[gphase]below:$a_1,b_1$}] (106) at (4.25, 1.75) {};
		\node [style=gn, label={[gphase]above:$a_2,b_2$}] (107) at (4.25, 0.25) {};
		\node [style=gn, label={[gphase]above:$a_n,b_n$}] (108) at (4.25, -1.75) {};
		\node [style=gn, label={[gphase]below:$c_1,d_1$}] (110) at (14.25, 1.75) {};
		\node [style=gn, label={[gphase]above:$c_2,d_2$}] (111) at (14.25, 0.25) {};
		\node [style=gn, label={[gphase]above:$c_n,d_n$}] (112) at (14.25, -1.75) {};
		\node [style=none] (113) at (0, -0.75) {\Fusion};
		\node [style=none] (114) at (0, 0.75) {\Colour};
	\end{pgfonlayer}
	\begin{pgfonlayer}{edgelayer}
		\draw (1.center) to (0);
		\draw (2.center) to (4);
		\draw (3.center) to (5);
		\draw (14.center) to (11);
		\draw (12.center) to (13);
		\draw (10.center) to (18);
		\draw (58) to (54.center);
		\draw (60) to (52.center);
		\draw (60) to (51.center);
		\draw (56) to (45.center);
		\draw (56) to (59.center);
		\draw (60) to (68.center);
		\draw (58) to (57.center);
		\draw (64) to (38.center);
		\draw (64) to (61.center);
		\draw (64) to (37.center);
		\draw (65) to (50.center);
		\draw (65) to (55.center);
		\draw (43) to (39.center);
		\draw (43) to (69.center);
		\draw (43) to (72.center);
		\draw (75) to (76);
		\draw (73) to (74);
		\draw (77) to (78);
		\draw [style=brace edge] (85.center) to (86.center);
		\draw (42.center) to (88);
		\draw (46.center) to (89);
		\draw (67.center) to (90);
		\draw (88) to (58);
		\draw (89) to (60);
		\draw (90) to (56);
		\draw (66.center) to (91);
		\draw (62.center) to (92);
		\draw (48.center) to (93);
		\draw (91) to (64);
		\draw (92) to (65);
		\draw (93) to (43);
		\draw (0) to (100);
		\draw [style={hadamard_edge}] (100) to (20.center);
		\draw [style={hadamard_edge}] (100) to (26.center);
		\draw (4) to (101);
		\draw [style={hadamard_edge}] (101) to (21.center);
		\draw [style={hadamard_edge}] (101) to (22.center);
		\draw [style={hadamard_edge}] (101) to (25.center);
		\draw [style={hadamard_edge}] (100) to (101);
		\draw (5) to (102);
		\draw [style={hadamard_edge}] (102) to (23.center);
		\draw [style={hadamard_edge}] (102) to (24.center);
		\draw [style={hadamard_edge}, bend right=45, looseness=0.75] (100) to (102);
		\draw (11) to (103);
		\draw [style={hadamard_edge}] (103) to (28.center);
		\draw [style={hadamard_edge}] (103) to (30.center);
		\draw [style={hadamard_edge}] (103) to (31.center);
		\draw (13) to (104);
		\draw [style={hadamard_edge}] (104) to (33.center);
		\draw [style={hadamard_edge}] (104) to (32.center);
		\draw (18) to (105);
		\draw [style={hadamard_edge}] (105) to (34.center);
		\draw [style={hadamard_edge}] (105) to (29.center);
		\draw [style={hadamard_edge}] (104) to (105);
	\end{pgfonlayer}
\end{tikzpicture}

%% file: conclusion.tex
\section{Conclusion}
\label{sec:conclusion}

We presented a simplified version of the qudit ZX-calculus for odd prime dimensions based on the work in Ref.\@~\cite{boothCompleteZXcalculiStabiliser2022}.
This version includes fewer rules and a new scalar gadget to bring the reasoning about scalars more in line with practice.
We also extended the spider-removing versions of local complementation and pivoting to qupits.
This extension enabled us to reduce diagrams efficiently to AP-form and its unique version, the reduced AP-form.
As a result, we obtained a new completeness proof for the qupit stabiliser fragment, which is more straightforward compared to previous proofs.
Additionally, we discovered a reduction to GSLC form, leading to a novel layered decomposition of qupit Clifford unitaries.
To support these developments, we implemented our rewrites into \texttt{DiZX},
a port of \texttt{PyZX} that now supports qudit stabiliser diagrams of arbitrary dimension.

For future work, it would be interesting to investigate whether our techniques can be applied to develop a useful circuit optimisation pipeline for qudits.
It would also be valuable to identify specific circuits that would benefit from such optimisation.

%% file: app-minimality.tex
\section{Necessity of the rules}

We can demonstrate that most of the non-scalar rules of our axiomatisation
are \emph{necessary}, meaning that they cannot be derived from the other rules.

Note that the standard approach to showing necessity involves defining an
alternative interpretation of the diagrams in which every rewrite rule remains
sound, except for the specific rule being examined for necessity.
This approach reveals that the other rules cannot establish the rule that
undermines soundness.
Several examples of this approach can be found in the works of Backens, Perdrix, and Wang~\cite{
  backensSimplifiedStabilizerZXcalculus2017,
  backensMinimalStabilizerZXcalculus2020}.
In particular, we may define an interpretation into projective Hilbert spaces
(quotienting by all non-zero scalars), in order to automatically satisfy all
the scalar axioms and focus on the non-scalar axioms.
This automatically satisfies all the scalar axioms, allowing us to focus
solely on the non-scalar axioms. Another approach involves using graph
properties that are invariant under all but one rule.
In the following discussion, we rely on the invariants of non-emptiness and
connectivity.

We can demonstrate the necessity of all but two of the stabiliser rules:
\begin{itemize}
  \item At least one of the \TextFusion rules is necessary, as they are the
  only rules that allow the decomposition of a spider with an arbitrary
  number of legs into spiders with fewer legs. In other words, these
  rules are not sound for an interpretation that assigns zero to all
  spiders with at least $p$ legs.
  \item \TextSpecial is the only axiom that enables the removal of all
  non-identity generators from a diagram. This breaks the interpretation
  where every generator is zero, except for the identity.
  \item \TextColour is necessary. To see this, consider the interpretation into
  projective Hilbert spaces where we redefine the X-spiders to swap the
  sign of the Pauli phase. It can be easily verified that this new
  interpretation satisfies all axioms except for \TextColour.
  \item \TextCopy is necessary since it is the only axiom that can transform a
  connected diagram into a disconnected one.
  \item \TextEuler is necessary, as shown by a modified interpretation similar
  to those in Refs.~\cite{duncanGraphStatesNecessity2009,
    gongEquivalenceLocalComplementation2017}.
\end{itemize}

We propose a conjecture regarding the necessity of \TextMElim, as it stands out
as the only rule that establishes a connection between elements in $\Z_p^*$ and
their multiplicative inverses. Although we lack a formal proof for this
intuition, we believe it to be true.

It is worth noting that despite its centrality
in most derivations, there remains one stabiliser axiom for
which we have no knowledge of its necessity, even in the qubit case~\cite{backensMinimalStabilizerZXcalculus2020}
or in the setting of graphical linear algebra~\cite{zanasiInteractingHopfAlgebras2018}: \TextBigebra.
We leave this intriguing open problem for the particularly motivated reader to
explore further.

As for the scalar rules, at least one subcase of each is necessary:
\begin{itemize}
  \item One subcase of \TextOmega is necessary because it is the only rule that
  allows the introduction of an $\omega$ scalar box, thereby breaking the
  interpretation where we redefine the $\omega$ scalar box.
  \item \TextZero is the only rule that relates a diagram without a zero scalar
  box to one that includes it. This means that when we interpret the zero
  scalar box as equal to 1 and set all other generators to zero, this rule
  becomes necessary.
  \item \TextOne is necessary as it is the only rule that connects a non-empty
  diagram to an empty diagram.
  \item \TextProd is necessary because there are complex numbers that cannot be
  expressed within the fragment of the language without scalar boxes. This rule
  is the only one that allows the multiplication of two such numbers.
  \item \TextNul is necessary, following an analogous argument as of
  Ref.~\cite{backensSimplifiedStabilizerZXcalculus2017}.
  \item In \TextGauss, the subcase $b=0$ is necessary since it is the only rule
  that allows one to interpret a diagram to a scalar box with a non-unit
  modulus. Additionally, at least one subcase $b \neq 0$ is necessary because
  these are the only rules that introduce a $-1$ scalar box.
\end{itemize}

%% file: app-lemmas.tex
\allowdisplaybreaks
\setlength{\jot}{20pt}

\section{Qupit Clifford ZX-calculus}

\subsection{Multipliers}

We extend our language by
\emph{multipliers}~\cite{boothCompleteZXcalculiStabiliser2022}, which are
defined as:
\begin{equation}
  \label{eq:mult-def}
  \input{definitions/multiplier.tikz}
  \qquad \qquad
  \input{definitions/multiplier-t.tikz}
\end{equation}
We can explicitly express multipliers as, for $x \in \Z_p^*$,
\begin{equation}
  \label{eq:multiplier_explicit}
  \input{equations/multiplier_explicit.tikz}
  \qquad \qquad
  \input{equations/multiplier_t_explicit.tikz}
\end{equation}
The following equations hold for multipliers and are proved in Ref
.\@~\cite{boothCompleteZXcalculiStabiliser2022}:
\begin{proposition}
  \label{prop:mult-prop}
  \begin{equation*}
    \input{equations/multiplier.tikz}
  \end{equation*}
\end{proposition}

\subsection{Recovering the derivations of
\texorpdfstring{Ref.~\cite{boothCompleteZXcalculiStabiliser2022}}{Booth and Carette, 2022}}

In this appendix, we recover all of the lemmas that were proved
in~\cite{boothCompleteZXcalculiStabiliser2022}. We do this by proving that
all of the axioms used there are derivable from the simplified set given in
section~\ref{sec:qupit-cliff} (up to scalars). We also show that the
language is complete for the scalar fragment, which completes the
proof. Since many of these proofs are entirely analogous to their
counterparts in Ref.~\cite{boothCompleteZXcalculiStabiliser2022}, we
omit them and refer to Ref.~\cite{boothCompleteZXcalculiStabiliser2022}
instead. In order to avoid ambiguity, we refer to the proofs in the specific
version of Ref.~\cite{boothCompleteZXcalculiStabiliser2022} cited as
Ref.~\cite{boothCompleteZXcalculiStabiliser2022v3}.

\begin{figure}[tbh]
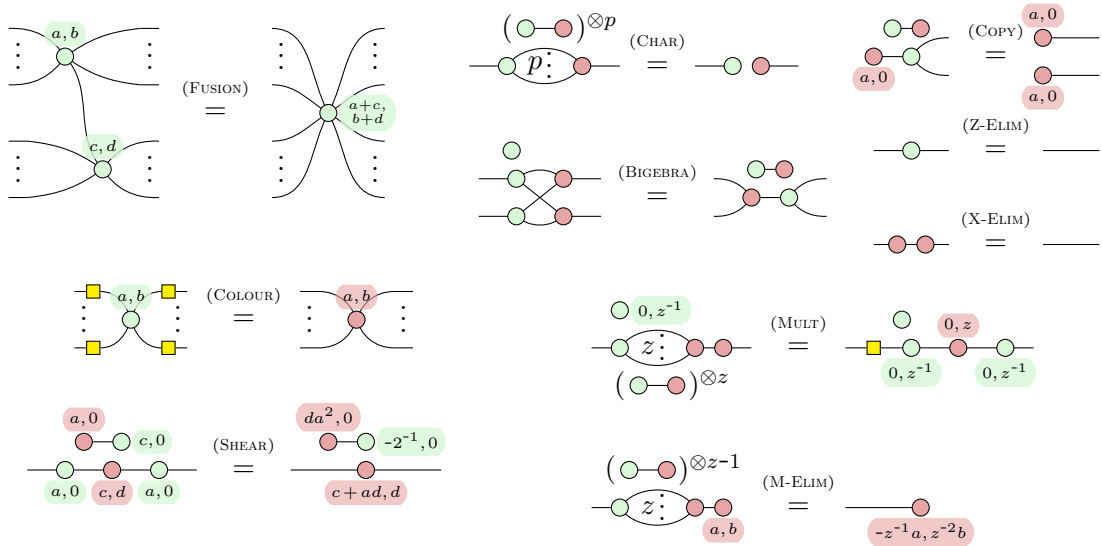

  \ctikzfig{figures/axioms-old}
  \caption{The original rule set of the qupit stabiliser ZX-calculus of~\cite{boothCompleteZXcalculiStabiliser2022}.}
  \label{fig:axioms-old}
\end{figure}

In~\cite{boothCompleteZXcalculiStabiliser2022}, the calculus was axiomatised
using the equations presented in Figure~\ref{fig:axioms-old}.

Comparing with the axioms of this paper (and ignoring scalars for now),
the missing axioms are \textsc{Z-Elim}, \textsc{X-Elim}, \textsc{Char},
\textsc{Mult}, \textsc{Shear}. In addition, the axioms of \textsc{Fusion}
and \textsc{Copy} were made more minimalistic.

\begin{lemma}
  \label{lem:green_trivial}
  Green \(1 \to 1\) spiders are trivial:
  \begin{equation*}
    \input{figures/equations/green_trivial.tikz} 
  \end{equation*}
\end{lemma}
\begin{lproof}
  \begin{equation*}
    \input{figures/equations/green_trivial_proof.tikz} \qedhere
  \end{equation*}
\end{lproof}

\begin{lemma}
  \label{lem:hadamard_product}
  Products of Hadamards are antipodes:
  \begin{equation*}
    \input{figures/equations/hadamard_product.tikz}
  \end{equation*}
\end{lemma}
\begin{lproof}
  Same as \booth{37}.
\end{lproof}

\begin{lemma}
  \label{lem:antipode_inverse}
  Antipodes are self-inverse:
  \begin{equation*}
    \input{figures/equations/antipode_inverse.tikz}
  \end{equation*}
\end{lemma}
\begin{lproof}
  \begin{equation*}
    \input{figures/equations/antipode_inverse_proof.tikz} \qedhere
  \end{equation*}
\end{lproof}

\begin{lemma}
  \label{lem:hadamard_antipode}
  Hadamards and antipodes commute:
  \begin{equation*}
    \input{figures/equations/hadamard_antipode.tikz}
  \end{equation*}
\end{lemma}
\begin{lproof}
  Same as \booth{38}.
\end{lproof}


\begin{lemma}
  \label{lem:hadamard_inverse}
  The inverse Hadamard is a product of Hadamards, and admits a ``tree''
  Euler decomposition:
  \begin{equation*}
    \input{figures/equations/hadamard-inverse.tikz}
  \end{equation*}
\end{lemma}
\begin{lproof}
  The first part is the same as \booth{39}, and the second part follows using
  \TextEuler and \TextColour.
  The last two follow from the first equation and \cref{lem:hadamard_product}.
\end{lproof}

\begin{lemma}
  \label{lem:hadamard_identity}
  The product of the Hadamard and its inverse equals the identity:
  \begin{equation*}
    \input{figures/equations/hadamard_identity.tikz}
  \end{equation*}
\end{lemma}
\begin{lproof}
  \begin{equation*}
    \input{figures/equations/hadamard_identity_proof.tikz}
  \end{equation*}
  The second equation can be proved similarly.
\end{lproof}

\begin{lemma}
  \label{lem:antipode_unit}
  Units absorb antipodes:
  \begin{equation*}
    \input{figures/equations/antipode_unit.tikz}
  \end{equation*}
\end{lemma}
\begin{lproof}
  Same as \booth{40}.
\end{lproof}

\begin{lemma}
  \label{lem:antipode_phase}
  For any \(x,y \in \Z_p\),
  \begin{equation*}
    \input{figures/equations/antipode_phase.tikz}
  \end{equation*}
\end{lemma}
\begin{lproof}
  \begin{align*}
    &\input{figures/equations/antipode_phase_proof_red.tikz} \\
    &\input{figures/equations/antipode_phase_proof_green.tikz} \qedhere
  \end{align*}
\end{lproof}

\begin{lemma}
  \label{lem:bigebra_mn}
  The bigebra law holds for multiple legs:
  for any \(m,n \in \N\), $2 \leq n,m$,
  \begin{equation*}
    \input{figures/equations/bigebra_arbitrary.tikz}\quad,
  \end{equation*}
  where in the diagram on the LHS, there are \(m\) green and \(n\) red spiders,
  and each green spider is connected to each red spider by a single wire.
\end{lemma}
\begin{lproof}
  Follows from straightforward induction (which furthermore is analogous to the qubit case).
\end{lproof}

\begin{lemma}
  \label{lem:copy_og}
  Using the \TextCopy rule in Figure~\ref{fig:axioms}, the \textsc{Copy} rule
  in Ref.~\cite{boothCompleteZXcalculiStabiliser2022} is derivable:
  \begin{equation*}
    \input{figures/equations/copy_og.tikz}
  \end{equation*}
\end{lemma}
\begin{lproof}
  The following holds for $a = 2,\cdots,p$.
  As $p \text{ mod } p = 0$, this also proves the $a = 0$ subcase of \TextCopy.
  The $a = 1$ case is just \TextCopy.
  \begin{align*}
    \input{figures/equations/copy_og_proof-0.tikz} \quad
    &\input{figures/equations/copy_og_proof-1.tikz} \\
    &\input{figures/equations/copy_og_proof-2.tikz} \qedhere
  \end{align*}
\end{lproof}

\begin{lemma}
  \label{lem:hopf}
  The Hopf identity is derivable in \(\ZXeq\):
  \begin{equation*}
    \input{figures/equations/hopf.tikz}
  \end{equation*}
\end{lemma}
\begin{lproof}
  \begin{equation*}
    \input{figures/equations/hopf_proof.tikz} \qedhere
  \end{equation*}
\end{lproof}

\begin{lemma}
  \label{lem:Char}
  The axiom \textsc{Char} of~\cite{boothCompleteZXcalculiStabiliser2022}
  is derivable:
  \begin{equation*}
    \input{figures/equations/Char.tikz}
  \end{equation*}
\end{lemma}
\begin{lproof}
  \begin{equation*}
    \input{figures/equations/Char_proof.tikz} \qedhere
  \end{equation*}
\end{lproof}

We are now ready to prove the completeness of the calculus for elementary
scalars.
\elementaryScalarCompleteness*
\begin{lproof}
  First, note that we have:
  \begin{equation*}
    \input{figures/equations/scalar_sqrt_p_inverse.tikz}
  \end{equation*}
  We can use this rule and the scalar axioms to rewrite every scalar in
  \cref{def:elementary-scalar}, as well as
  the zero scalar \input{figures/equations/zero_diagram.tikz}, into an explicit
  scalar. Then, we apply \TextProd to rewrite this collection of explicit
  scalars into a single one.
\end{lproof}


\begin{lemma}
  \label{lem:loop}
  Self-loops on green spiders can be eliminated:
  \begin{equation*}
    \input{figures/equations/loop.tikz}
  \end{equation*}
  We include the colour-swapped version of this rule for completeness, even
  though it no longer includes a genuine self-loop.
\end{lemma}
\begin{lproof}
  \begin{equation*}
    \input{figures/equations/loop_proof.tikz}
  \end{equation*}
  The red version follows form \TextColour and \cref{lem:hadamard_product}.
\end{lproof}


\begin{lemma}
  \label{lem:unit_rotation_elim}
  Green units absorb red rotations and vice-versa:
  \begin{equation*}
    \input{figures/equations/unit_rotation_elim.tikz}
  \end{equation*}
\end{lemma}
\begin{lproof}
  \begin{equation*}
    \input{figures/equations/unit_rotation_elim_proof_green.tikz}
  \end{equation*}
  The red version follows form \TextColour.
\end{lproof}

\begin{lemma}
  \label{lem:antipode_copy}
  The green co-multiplication copies antipodes:
  \begin{equation*}
    \input{figures/equations/antipode_copy.tikz}
  \end{equation*}
\end{lemma}
\begin{lproof}
  \begin{equation*}
    \input{figures/equations/antipode_copy_proof.tikz} \qedhere
  \end{equation*}
\end{lproof}

\begin{lemma}
  \label{lem:push_pauli_state}
  Green Pauli states copy through red rotations and vice-versa:
  \begin{equation*}
    \input{new/pauli-state-push.tikz}
  \end{equation*}
\end{lemma}
\begin{lproof}
  \begin{align*}
    &\input{new/pauli-state-push-proof.tikz} \\
    &\input{new/pauli-state-push-compl-proof.tikz} \qedhere
  \end{align*}
\end{lproof}

\begin{lemma}
  \label{lem:antipode_multiplier}
  The antipode can be rewritten as a multiplication:
  \begin{equation*}
    \input{figures/equations/antipode_multiplier.tikz}
  \end{equation*}
\end{lemma}
\begin{lproof}
  This follows from \TextSpecial and the definition of the multiplier, \cref{eq:mult-def}.
\end{lproof}

\begin{lemma}
  \label{lem:antipode_spider}
  For any \(x,y \in \Z_p\) and \(m,n \in \N\),
  \begin{equation*}
    \input{figures/equations/antipode_spider.tikz}
  \end{equation*}
\end{lemma}
\begin{lproof}
  \begin{align*}
    &\input{figures/equations/antipode_spider_proof.tikz} \\
    &\input{figures/equations/antipode_spider_proof_compl.tikz} \qedhere
  \end{align*}
\end{lproof}

\begin{lemma}
  \label{lem:colour}
  We can derive the \TextColour rule for both red and green spiders, and
  also for the Hadamard inverse:
  \begin{align*}
    \input{figures/equations/colour_commutation_green.tikz}
    \qquad & \qquad
    \input{figures/equations/colour_commutation_red.tikz}
    \\
    \input{figures/equations/colour_commutation_green_minu.tikz}
    \qquad & \qquad
    \input{figures/equations/colour_commutation_red_minu.tikz}
  \end{align*}
\end{lemma}
\begin{lproof}
  This follows from \TextColour and
  \cref{lem:hadamard_inverse,lem:antipode_spider}.
\end{lproof}

\begin{lemma}
  \label{lem:hadamard_push}
  Hadamard gates or their inverses can be pushed through spiders:
  \begin{equation*}
    \input{figures/equations/colour_commutation.tikz}
  \end{equation*}
\end{lemma}
\begin{lproof}
  This follows from \cref{lem:colour,lem:hadamard_identity}.
\end{lproof}

\begin{lemma}
  \label{lem:pauli_copy_phase}
  Green spiders copy red Pauli phases, and vice-versa: for any \(x \in \Z_p\),
  \begin{equation*}
    \input{figures/equations/pauli_copy_phase.tikz}
  \end{equation*}
\end{lemma}
\begin{lproof}
  \begin{equation*}
    \input{figures/equations/pauli_copy_phase_proof.tikz}
  \end{equation*}
  The second equation follows from \cref{lem:colour,lem:hadamard_identity}.
\end{lproof}

\begin{lemma}
  \label{lem:multiplier_sum}
  Parallel multipliers sum: for any \(x,y \in \Z_p\):
  \begin{equation*}
    \input{figures/equations/multiplier_sum.tikz}
  \end{equation*}
\end{lemma}
\begin{lproof}
  This is a straightforward consequence of \TextFusion.
\end{lproof}

\begin{lemma}
  \label{lem:multiplier_elim}
  For any \(z \in \Z_p^*\),
  \begin{equation*}
    \input{figures/equations/multiplier_elim.tikz}
  \end{equation*}
\end{lemma}
\begin{lproof}
  \begin{equation*}
    \input{figures/equations/multiplier_elim_proof_green.tikz}
  \end{equation*}
  The other rule follows from \TextColour.
\end{lproof}

\begin{lemma}
  \label{lem:multiplier_product}
  For any \(x,y \in \N\),
  \begin{equation*}
    \input{figures/equations/multiplier_product.tikz}
  \end{equation*}
\end{lemma}
\begin{lproof}
  \begin{equation*}
    \input{figures/equations/multiplier_product_proof.tikz}
  \end{equation*}
  The second equality follows from \TextColour.
\end{lproof}

\begin{lemma}
  \label{lem:multiplier_inverse}
  For any \(x \in \Z_p^*\),
  \begin{equation*}
    \input{figures/equations/multiplier_inverse.tikz}
  \end{equation*}
\end{lemma}
\begin{lproof}
  \begin{equation*}
    \input{figures/equations/multiplier_inverse_proof.tikz}
  \end{equation*}
  The second equality follows from \TextColour.
\end{lproof}

\begin{lemma}
  \label{lem:multiplier_copy}
  Spiders copy invertible multipliers: for any \(x \in \Z_p^*\),
  \begin{equation*}
    \input{figures/equations/multiplier_copy.tikz}
  \end{equation*}
\end{lemma}
\begin{lproof}
  \begin{equation*}
    \input{figures/equations/multiplier_copy_proof_green.tikz}
  \end{equation*}
  The other equation follows form \TextColour and the definition of the multiplier, \cref{eq:mult-def}.
\end{lproof}

\begin{lemma}
  \label{lem:multiplier_spider}
  The action of multipliers on spiders is given by, for any \(x \in \Z_p^*\),
  \begin{equation*}
    \input{figures/equations/multiplier_spider.tikz}
  \end{equation*}
\end{lemma}
\begin{lproof}
  This follows from \cref{lem:multiplier_copy} and \TextMElim.
\end{lproof}

\begin{restatable}{lemma}{pushmult}
  \label{lem:multiplier_push}
  We can \enquote{push} multipliers through spiders as follows, for any $a, b
  \in \Z_p$ and $x \in \Z_p^*$,
  \begin{equation*}
    \input{new/push-mult-green.tikz}
    \qquad \qquad
    \input{new/push-mult-inv-green.tikz}
  \end{equation*}
  \begin{equation}
    \input{new/push-mult-red.tikz}
    \qquad \qquad
    \input{new/push-mult-inv-red.tikz}
  \end{equation}
\end{restatable}
\begin{lproof}
  \begin{align*}
      &\input{new/push-mult-green-proof.tikz}\\
      &\input{new/push-mult-red-proof.tikz}
  \end{align*}
  The other proofs follow from the above equations while using the
  multiplicative inverse of the multipliers.
\end{lproof}

\begin{lemma}
  \label{lem:multiplier_had}
  The product of a multiplier and a Hadamard gate is an H-box:
  \begin{equation*}
    \input{equations/hadamard_multiplier.tikz}
  \end{equation*}
\end{lemma}
\begin{lproof}
  This follows from \cref{lem:hadamard_identity} and the definition of the multiplier, \cref{eq:mult-def}.
\end{lproof}

\prophbox*
\begin{lproof}
  The bottom two equations follow from the definition of the H-box at \cref{eq:hbox_def} and \cref{lem:green_trivial}.
  The rest can be proved using \cref{lem:multiplier_had,prop:mult-prop,lem:hadamard_push,lem:hadamard_identity}.
\end{lproof}

\begin{restatable}{lemma}{xhadmult}
  \label{lem:multiplier_hbox}
  H-boxes multiply with multipliers, for any $x, y \in \Z_p$,
  \begin{equation*}
    \input{new/xhad-mult.tikz}
  \end{equation*}
\end{restatable}
\begin{proof}
  \begin{equation*}
    \input{new/xhad-mult-proof-1.tikz}
  \end{equation*}
  \begin{equation*}
    \input{new/xhad-mult-proof-2.tikz} \qedhere
  \end{equation*}
\end{proof}

\begin{restatable}{lemma}{pushxhad}
  \label{lem:hbox_push}
  We can \enquote{push} H-boxes through spiders as follows, for any $a, b \in
  \Z_p$ and $x \in \Z_p^*$,
  \begin{equation*}
    \input{new/push-xhad-green.tikz}
    \qquad \qquad
    \input{new/push-xhad-red.tikz}
  \end{equation*}
\end{restatable}
\begin{lproof}
  First of all,
  \begin{equation*}
    \input{new/push-xhad-green-proof.tikz}
  \end{equation*}
  The other equation can be proved similarly,
  \begin{equation*}
    \input{new/push-xhad-red-proof.tikz} \qedhere
  \end{equation*}
\end{lproof}

\begin{lemma}
  \label{lem:scalar-xxhad}
  \begin{equation*}
    \input{new/scalar_xxhad.tikz}
  \end{equation*}
\end{lemma}
\begin{lproof}
  \begin{equation*}
    \input{new/scalar_xxhad_proof.tikz} \qedhere
  \end{equation*}
\end{lproof}

\begin{lemma}
  \label{lem:clifford_states}
  Any purely-Clifford states can be represented in both the red and green
  fragment: for any \(x \in \Z_p^*\),
  \begin{equation*}
    \input{figures/equations/clifford_states.tikz}
  \end{equation*}
\end{lemma}
\begin{lproof}
  Firstly, we prove the subcase \(x=1\) of (a):
  \begin{equation*}
    \input{figures/equations/clifford_states_proof_green.tikz}
  \end{equation*}
  so that
  \begin{equation*}
    \input{figures/equations/clifford_states_proof_green_2.tikz}
  \end{equation*}
  and
  \begin{equation*}
    \input{figures/equations/clifford_states_proof_green_3.tikz}.
  \end{equation*}
  Then the general case for any invertible \(x\) follows using
  lemma~\ref{lem:multiplier_spider}.

  (b) follows once again using \TextColour.
\end{lproof}

\begin{lemma}
  \label{lem:hadamard_euler}
  The Hadamards admit more standard Euler decompositions (originally shown
  for qudit ZX in~\cite{wangQufiniteZXcalculusUnified2022}):
  \begin{equation*}
    \input{figures/equations/hadamard_euler.tikz}
  \end{equation*}
\end{lemma}
\begin{lproof}
  \begin{equation*}
    \input{figures/equations/hadamard_euler_proof.tikz}
  \end{equation*}
  We obtain the second derivation, as always, using \TextColour.
\end{lproof}

In the next few proofs, we make frequent use of the following fact:
\begin{lemma}
  \label{lem:sum_of_squares}
  For any \(x \in \Z_p\), there are \(a,b \in \Z_p\) such that \(x = a^2
  + b^2\).
\end{lemma}
\begin{lproof}
  This is true in general for any finite field.
  See~\cite{yutsumuraEachElementFinite2017} for a proof.
\end{lproof}

\begin{lemma}
  \label{lem:euler_squares}
  For any \(z \in \Z_p^*\),
  \begin{equation*}
    \input{figures/euler/euler_squares.tikz}
  \end{equation*}
\end{lemma}
\begin{lproof}
  We have
  \begin{equation*}
    \input{figures/euler/euler_colour_swap_proof.tikz}
  \end{equation*}
  so that
  \begin{equation*}
    \input{figures/euler/euler_squares_proof.tikz}
  \end{equation*}
  and
  \begin{equation*}
    \input{figures/euler/euler_squares_proof_2.tikz}
  \end{equation*}
  The second version is obtained using a completely analogous argument.
\end{lproof}

\begin{lemma}
  \label{lem:euler_tree}
  For any \(z \in \Z_p^*\) (not just squares),
  \begin{equation*}
    \input{figures/euler/euler_tree.tikz}
  \end{equation*}
\end{lemma}
\begin{lproof}
  If \(z\) is a square, then this result is immediate by the previous lemma. Otherwise, by Lemma~\ref{lem:sum_of_squares}, \(z =
  a^2 + b^2\) and \(a,b \in \Z_p\) are non-zero. Then
  \begin{equation*}
    \input{figures/euler/euler_tree_proof.tikz} \qedhere
  \end{equation*}
\end{lproof}

\begin{lemma}
  \label{lem:H_loop}
  Hadamard loops correspond to purely-Clifford operations: for any \(x \in \Z_p\)
  and \(z \in \Z_p^*\),
  \begin{equation*}
    \input{figures/equations/H_loop.tikz}
  \end{equation*}
\end{lemma}
\begin{lproof}
  The case \(x=0\) is clear by decomposing the H-box according to \cref{eq:hbox_def}.
  Therefore, we only need to show that for \(z \in \Z_p^*\):
  \begin{align*}
    \input{figures/equations/H_loop_proof_green-0.tikz} \quad
    &\input{figures/equations/H_loop_proof_green-1.tikz} \\
    &\input{figures/equations/H_loop_proof_green-2.tikz}
  \end{align*}
  Under the assumption that the weight is invertible, the red version follows
  using \TextColour and the green version:
  \begin{equation*}
    \input{figures/equations/H_loop_proof_red.tikz}
  \end{equation*}
\end{lproof}

\lemSupplementarity*
\begin{lproof}
  If \(b = x^2\) for some non-zero \(x \in \Z_p\), then
  \begin{equation*}
    \input{figures/euler/fork_identity_proof.tikz}
  \end{equation*}
  Otherwise \(b = s^2 + t^2\) where \(s,t \in \Z_p\) are non-zero, and
  \begin{equation*}
    \input{figures/euler/fork_identity_proof_2.tikz}
  \end{equation*}
  The second equation follows from the first equation and the application of \TextColour.
\end{lproof}

\lemStrictCliffordColour*
\begin{lproof}
  Firstly,
  \begin{equation*}
    \input{figures/euler/state_colour_change_proof_1.tikz}
  \end{equation*}
  so that
  \begin{equation*}
    \input{figures/euler/state_colour_change_proof_2.tikz}
  \end{equation*}
  whence
  \begin{equation*}
    \input{figures/euler/state_colour_change_proof_3.tikz}
  \end{equation*}
  and, finally,
  \begin{equation*}
    \input{figures/euler/state_colour_change_proof_4.tikz}
  \end{equation*}
  The change of colour in the scalar, as well as the second derivation,
  follow using \TextColour.
\end{lproof}

\begin{lemma}
  \label{lem:scalar_z_equiv}
  The following states are equivalent:
  \begin{equation*}
    \input{equations/scalar_minuzminu1_z.tikz}
  \end{equation*}
\end{lemma}
\begin{lproof}
  \begin{equation*}
    \input{equations/scalar_minuzminu1_z_proof.tikz} \qedhere
  \end{equation*}
\end{lproof}


\begin{lemma}
  \label{lem:pauli_push}
  X-spiders with arbitrary phases copy Pauli Z-spiders and vice-versa:
  \begin{equation*}
    \input{new/extended-copy.tikz}
    \qquad \qquad
    \input{new/extended-copy-compl.tikz}
  \end{equation*}
\end{lemma}
\begin{lproof}
  First of all,
  \begin{equation*}
    \input{new/extended-copy-proof-1.tikz}
  \end{equation*}
  Then, we separate the equation into two cases based on whether the Z-spider
  is Pauli or not.
  In case $d = 0$, the Z-spider is Pauli and therefore:
  \begin{equation*}
    \input{new/extended-copy-proof-2.tikz}
  \end{equation*}
  Note that if $d = 0$, then $ad - c = -c$ and so the lemma holds.
  Otherwise, $d \neq 0$ and therefore $d^{\minu 1}$ exists, so we can apply
  the state-change lemma:
  \begin{equation*}
    \input{new/extended-copy-proof-3.tikz}
  \end{equation*}
  Note that the phases after the application of the second state-change
  follow from:
  \begin{equation*}
    -(a - c d^{\minu 1})(\minu d^{\minu 1})^{\minu 1},
    \,  \minu(\minu d^{\minu 1})^{\minu 1}
    = -(a - c d^{\minu 1}) (\minu d), \,  d
    = ad - c, \, d
  \end{equation*}
  We can prove the second equation of the lemma using Hadamard-boxes as
  follows:
  \begin{equation*}
    \input{new/extended-copy-compl-proof.tikz} \qedhere
  \end{equation*}
\end{lproof}

We are now ready to prove that axioms \textsc{Mult} and \textsc{Shear}
of~\cite{boothCompleteZXcalculiStabiliser2022} are derivable from our
simplified set of axioms:
\begin{proposition}
  \label{prop:mult}
  For any \(z \in \Z_p^*\),
  \begin{equation*}
    \input{figures/euler/mult.tikz}
  \end{equation*}
\end{proposition}
\begin{lproof}
  We have
  \begin{equation*}
    \input{figures/euler/mult_finally.tikz}
  \end{equation*}
  so that
  \begin{equation*}
    \input{figures/euler/mult_finally_2.tikz} \qedhere
  \end{equation*}
\end{lproof}

\begin{proposition}
  For any \(a,c,d \in \Z_p\),
  \begin{equation*}
    \input{figures/equations/shear.tikz}
  \end{equation*}
\end{proposition}
\begin{lproof}
  We first prove the subcase \(c = 0\) and $d \neq 0$:
  \begin{equation*}
    \input{figures/equations/shear_proof_1.tikz}
  \end{equation*}
  Now, we have
  \begin{equation*}
    \input{figures/equations/shear_proof_2.tikz}
  \end{equation*}
  so that
  \begin{equation*}
    \input{figures/equations/shear_proof_3.tikz}
  \end{equation*}
  Now, if \(d = 0\),
  \begin{equation*}
    \input{figures/equations/shear_proof_4.tikz}
  \end{equation*}
  and putting both of these derivations together:
  \begin{equation*}
    \input{figures/equations/shear_proof_5.tikz} \qedhere
  \end{equation*}
\end{lproof}


\subsection{Graph-like diagrams}\label{app:graph-like}

\begin{proposition}
  \label{prop:local_complementation}
  $\gamma$-weighted local $\Z_d$-complementation is derivable in $\ZXeq$,
  for any graph $G = (V,E)$, $\gamma \in \Z_p$ and $u \in V$,
  \begin{equation}
    \label{eq:graph-lc}
    \input{figures/new/graph-local-compl.tikz}
  \end{equation}
\end{proposition}
\begin{lproof}
  Same as \booth{12}.
\end{lproof}

\subsubsection{Local complementation simplification}

\lc*
\begin{lproof}
  First, we can prove a simplified version of the lemma without phases of
  the boundary spiders and H-edges as follows:
  \begin{align*}
    &\input{new/local_compl_phaseless_proof-1.tikz} \\
    &\input{new/local_compl_phaseless_proof-2.tikz} \\
    &\input{new/local_compl_phaseless_proof-3.tikz} \\
    &\input{new/local_compl_phaseless_proof-4.tikz}
  \end{align*}
  Then, we can use the previous equation to prove the lemma.
  \begin{align*}
    &\input{new/local_compl_proof-1.tikz} \\
    &\quad \input{new/local_compl_proof-2.tikz} \\
    &\input{new/local_compl_proof-3.tikz} \qedhere
  \end{align*}
\end{lproof}

\subsubsection{Pivoting simplification}

\pivotpartial*
\begin{lproof}
  First, we can prove a simplified version of the equation that omits the
  phases of boundary spiders as follows,
  \begin{align*}
    &\input{new/partial_pivot_proof-1.tikz} \\
    &\input{new/partial_pivot_proof-2.tikz} \\
    &\input{new/partial_pivot_proof-3.tikz} \\
    &\input{new/partial_pivot_proof-4.tikz} \\
    &\input{new/partial_pivot_proof-5.tikz}
  \end{align*}
  Then, we can use the previous equation to prove the lemma as follows:
  \begin{align*}
    &\input{new/partial_pivot_phase_proof-1.tikz} \\
    &\input{new/partial_pivot_phase_proof-2.tikz} \qedhere
  \end{align*}
\end{lproof}

Now, we prove the general version of pivoting.
\pivot*
\begin{lproof}
  \begin{align*}
    &\input{new/pivot_proof-1.tikz} \\
    &\input{new/pivot_proof-2.tikz} \\
    &\input{new/pivot_proof-3.tikz} \\
    &\input{new/pivot_proof-4.tikz} \qedhere
  \end{align*}
\end{lproof}

\section{A normal form}

\apform*
\begin{lproof}
  We can prove this claim purely diagrammatically, by composing the diagram
  of \cref{eq:ap-form} with an effect that corresponds to the vector $\bra x$.
  By rewriting the diagram while keeping track of the scalars, we can prove
  that the diagram indeed represents the one described in
  \cref{eq:ap-form-poly}.
  These transformations are as follows:
  \begin{align*}
    &\input{figures/new/ap_form_proof-1.tikz} \\
    &\input{figures/new/ap_form_proof-2.tikz} \\
    &\input{figures/new/ap_form_proof-3.tikz}
  \end{align*}
  Note that if a Z-spider with no legs has phase $(z,0)$ for any $z \in \Z_p
  ^*$, then it equals the zero scalar.
  This means that the probability of such an effect is $0$.
  Therefore, the above diagram allows only such $\vec x$ vectors that
  satisfy the equation $E \vec x = \vec a$.
  Furthermore, the scalars that are copied from the phases part of the
  diagram equal the $\omega^{\phi(\vec x)}$ component of the equation.
  We conclude that a diagram in \cref{eq:ap-form} indeed equals the state
  presented in \cref{eq:ap-form-poly}.
\end{lproof}

\subsection{Completeness}

\apunique*
\begin{lproof}
  Since $\ket \psi \neq 0$, the set $\mathcal{A} = \{\vec x \mid E \vec{x} = \vec{a}\}$ is non-empty.
  Therefore, there is a unique system of equations in RREF that define $\mathcal{A}$.
  This means that $E$ and $\vec a$ are uniquely fixed.
  Now, for any assignment $\{x_{i_1} \coloneqq c_1,\, \ldots\, , x_{i_k} \coloneqq c_k\}$ of free variables, there exists a state $\ket{\vec x} \in \mathcal{A}$ such that $x_{i_\mu} = c_\mu$.
  Therefore, we have $\ip{\vec x}{\psi} = \omega^{\phi(c_1,\, \ldots\, , c_k)}$ for some fixed constant $\lambda \neq 0$.
  Using this fact we can determine the value of $\phi$ at all inputs $(c_1,\, \ldots\, , c_k)$ which is enough to compute each coefficient of $\phi$.
  We conclude that $\phi$ is uniquely fixed by $\ket \psi$.
\end{lproof}

\rowadd*
\begin{lproof}
  Firstly, we show that we can transform two disconnected X-states:
  \begin{align*}
    &\input{figures/new/row_add_sub_lem_proof_new-1.tikz} \\
    &\input{figures/new/row_add_sub_lem_proof_new-2.tikz}
  \end{align*}
  Then, we can show that we can transform a diagram in AP-form as follows:
  \begin{align*}
      &\input{figures/new/row_add_proof-1.tikz} \\
      &\input{figures/new/row_add_proof-2.tikz} \qedhere
  \end{align*}
\end{lproof}

\begin{restatable}{lemma}{removepauli}
  \label{lem:remove_pauli}
  We can remove Pauli-phases from the pivot spiders of diagrams in AP-form.
\end{restatable}
\begin{lproof}
  For any $a, x, e_i \in \Z_p$ where $i \in \{2, \ldots, k\}$ and $e_1 \in \Z_p^*$:
  \begin{equation*}
    \input{figures/new/remove_pauli_node.tikz}
  \end{equation*}
  where $a' \coloneqq -(a + x e_1^{\minu 1})$.
\end{lproof}

\begin{restatable}{lemma}{removeclif}
  \label{lem:remove-clif}
  We can remove strictly-Clifford phases from the pivot spiders of diagrams
  in AP-form.
\end{restatable}
\begin{lproof}
  To prove this case, we first show that we can push strictly-Clifford Z-spider
  through an X-spider with weighted outputs.
  That is, for any $a, e_i \in \Z_p$ where $i \in \{1, \ldots, k\}$ and $z \in \Z_p^*$:
  \begin{equation*}
    \input{figures/new/remove_cliff_node_sub_lem_proof.tikz}
  \end{equation*}
  Therefore, for any $a, x, e_i \in \Z_p$ where $i \in \{2, \ldots, k\}$ and $z,e_1 \in \Z_p^*$:
  \begin{align*}
    &\input{figures/new/remove_cliff_node-1.tikz} \\
    &\input{figures/new/remove_cliff_node-2.tikz}
  \end{align*}
  where
  $A_i = (a z e^{\minu 1} - x) e^{\minu 1} e_i$,
  $B_i = z e_1^{\minu 2} e_i^{2}$, and
  $E_{i,j} = z e_1^{\minu 2} e_i e_j$,
\end{lproof}

\begin{restatable}{lemma}{remhadedge}
  \label{lem:rem_had_edge}
  We can remove an H-edge between the pivot spider and a boundary spider
  that connects to the same internal spider as the pivot.
\end{restatable}
\begin{lproof}
  Let us suppose that the pivot spider is connected to the $\ell$-th wire
  with an H-box.
  Then, for any $a, x, e_i \in \Z_p$ where $i \in \{2, \ldots, k\}$ and $e_1 \in \Z_p^*$:
  \begin{equation*}
    \input{figures/new/remove_had_edge.tikz} \qedhere
  \end{equation*}
\end{lproof}

\begin{restatable}{lemma}{remhadedge2}
  \label{lem:rem_had_edge2}
  We can remove an H-edge between the pivot spider and a boundary spider
  that does not connect to the same internal spider as the pivot.
\end{restatable}
\begin{lproof}
  For any $a, b, x, e_i, f_h \in \Z_p$ where $i \in \{2, \ldots, k\}$, $h \in \{1, \ldots, j\}$ and $e_1 \in \Z_p^*$:
  \begin{equation*}
    \input{figures/new/remove_had_edge_sep.tikz} \qedhere
  \end{equation*}
\end{lproof}

\reducedap*
\begin{lproof}
  First, we can convert any diagram in $\ZXp$ into one in AP-form using local
  complementation and pivoting. Then, we can translate such a diagram into
  AP-form with a biadjacency matrix in RREF using Gaussian elimination, as
  demonstrated in \cref{lem:row-add}. Furthermore, we have established the
  proofs for removing any phase from the pivot spider (\cref{lem:remove_pauli}
  and \cref{lem:remove-clif}), as well as removing any H-edge connected to the
  pivot spider (\cref{lem:rem_had_edge} and \cref{lem:rem_had_edge2}).
  These results allow us to transform a diagram in such a way that its phase
  function $\phi$ only contains free variables from the equation system
  $E \vec x = \vec a$. Consequently, we can conclude that any diagram in
  $\ZXp$ can be rewritten into a form that satisfies the necessary properties
  to be considered a diagram in reduced AP-form.
\end{lproof}

\completeness*
\begin{lproof}
Without loss of generality, we can assume that both $A$ and $B$ are states by
map-state duality.
If $A$ and $B$ represent the same linear map, i.e.\@
$\interp{A} = \interp{B}$, then their reduced AP-forms are identical,
thanks to the uniqueness of the form proved in \cref{lem:ap-unique}.
Therefore, we can transform both $A$ and $B$ into diagrams in reduced AP-form
using \cref{lem:reduced-ap}. The sequence of transformations from $A$ to $A$ in reduced AP-form,
composed with the series of rewrites from $B$ in reduced AP-form to $B$,
provides us with a sequence of rewrites that transforms $A$ into $B$\@.
\end{lproof}

\boundarypivot*
\begin{lproof}
  Unfuse spiders and introduce Hadamards as follows:
  \begin{align*}
    &\input{new/partial-pivot-boundary-pf-1.tikz} \\
    &\input{new/partial-pivot-boundary-pf-2.tikz}
  \end{align*}
  Where in the last step we applied the regular pivot Lemma~\ref{lem:pivot-partial}.
\end{lproof}

%% file: figures/definitions/multiplier.tikz
\begin{tikzpicture}
	\begin{pgfonlayer}{nodelayer}
		\node [style=had] (0) at (2.5, 0) {$x$};
		\node [style=none] (1) at (1.25, 0) {};
		\node [style=none] (3) at (0, 0) {$\coloneqq$};
		\node [style=none] (4) at (-1.25, 0) {};
		\node [style=rmat] (5) at (-2.5, 0) {$x$};
		\node [style=none] (6) at (-3.75, 0) {};
		\node [style=had] (7) at (3.75, 0) {$\minu$};
		\node [style=none] (8) at (5, 0) {};
	\end{pgfonlayer}
	\begin{pgfonlayer}{edgelayer}
		\draw (1.center) to (0);
		\draw (5) to (4.center);
		\draw (6.center) to (5);
		\draw (0) to (7);
		\draw (8.center) to (7);
	\end{pgfonlayer}
\end{tikzpicture}

%% file: figures/definitions/multiplier-t.tikz
\begin{tikzpicture}
	\begin{pgfonlayer}{nodelayer}
		\node [style=had] (0) at (3.75, 0) {$x$};
		\node [style=none] (1) at (1.25, 0) {};
		\node [style=none] (2) at (0, 0) {$\coloneqq$};
		\node [style=none] (3) at (-1.25, 0) {};
		\node [style=lmat] (4) at (-2.5, 0) {$x$};
		\node [style=none] (5) at (-3.75, 0) {};
		\node [style=had] (6) at (2.5, 0) {$\minu$};
		\node [style=none] (7) at (5, 0) {};
	\end{pgfonlayer}
	\begin{pgfonlayer}{edgelayer}
		\draw (4) to (3.center);
		\draw (5.center) to (4);
		\draw (1.center) to (6);
		\draw (6) to (0);
		\draw (0) to (7.center);
	\end{pgfonlayer}
\end{tikzpicture}

%% file: figures/equations/zero_diagram.tikz
\begin{tikzpicture}
    \begin{pgfonlayer}{nodelayer}
        \node [style=gn, label={[gphase]left:$1,0$}] (0) at (0, 0) {};
    \end{pgfonlayer}
\end{tikzpicture}

%% file: figures/equations/H_loop_proof_green-0.tikz
\begin{tikzpicture}
	\begin{pgfonlayer}{nodelayer}
		\node [style=gn] (0) at (0, 0) {};
		\node [style=none] (1) at (1, 0) {};
		\node [style=had] (2) at (-1, 0) {$z$};
	\end{pgfonlayer}
	\begin{pgfonlayer}{edgelayer}
		\draw (0) to (1.center);
		\draw [bend left=90, looseness=1.50] (2) to (0);
		\draw [bend left=90, looseness=1.50] (0) to (2);
	\end{pgfonlayer}
\end{tikzpicture}

%% file: figures/equations/scalar_minuzminu1_z.tikz
\begin{tikzpicture}
	\begin{pgfonlayer}{nodelayer}
		\node [style=gn, label={[gphase]below:$0,z^{\minu 1}$}] (0) at (1.5, 0) {};
		\node [style=gn, label={[gphase]below:$0,z$}] (26) at (-1.25, 0) {};
		\node [style=none] (27) at (0, 0) {$=$};
	\end{pgfonlayer}
\end{tikzpicture}